\documentclass[journal, comsoc]{IEEEtran}

\usepackage{hyperref}
\usepackage{multirow}
\usepackage{makecell}
\usepackage{mathtools}
\usepackage{booktabs}
\hypersetup{draft}
\usepackage{tikzsymbols}
\usepackage{mathpazo}
\usepackage[T1]{fontenc}
\usepackage{algorithmic}
\usepackage{pifont}
\usepackage{amsfonts}
\usepackage{graphicx}
\newtheorem{theorem}{Theorem}
\usepackage{subfig}
\usepackage{cite}
\usepackage{graphicx}
\usepackage[linesnumbered,ruled,vlined]{algorithm2e}
\usepackage{amsmath}               
\newtheorem{assumption}{Assumption}
\newtheorem{remark}{Remark}
\newtheorem{corollary}{Corollary}
\usepackage{enumitem}
\setlist[itemize]{leftmargin=*}

\begin{document}

\title{Exploiting Semantic Localization in Highly Dynamic Wireless Networks using Deep Homoscedastic Domain Adaptation}

\author{Lei~Chu,~\IEEEmembership{Senior Member,~IEEE,}
        Abdullah~Alghafis, and Andreas~F.~Molisch,~\IEEEmembership{Fellow,~IEEE}
\thanks{Lei Chu and Andreas F. Molisch are with the Ming Hsieh Department of Electrical Engineering, University of Southern California, Los Angeles, CA 90089, USA. Abdullah~Alghafis is with  Information Technology Research Institute, King Abdulaziz City for Science and Technology, Saudi Arabia.  }
\thanks{This work was supported partly by the National Science Foundation under project CIF 2008443 and partly by the Department of Transportation under the METRANS framework.}
\thanks{Part of this work was presented at IEEE ICC 2023\cite{chu2023mda}.}
}

\markboth{Journal of \LaTeX\ Class Files, ~Vol.~XX, No.~XX, December ~2022}%
{Shell \MakeLowercase{\textit{et al.}}: Bare Demo of IEEEtran.cls for Journals}

\maketitle

\begin{abstract}

\textcolor{black}{This research paper delves into leveraging Machine Learning (ML) for precise localization in GPS-challenged environments like urban canyons, addressing the complexities of time-varying signal propagation types, where transient obstructions, such as vehicles, can modify the channel state information (CSI) over time. It presents a novel approach termed semantic localization, which recognizes signal propagation conditions as semantic elements, incorporating them into the localization framework to enhance both accuracy and resilience. To tackle the issue of diverse CSIs at each location and the extensive need for labeled data, the paper proposes a multi-task deep domain adaptation (DA) strategy. This approach trains neural networks using a limited set of labeled data complemented by a vast array of unlabeled samples, coupled with innovative scenario adaptive learning techniques for optimal representation learning and knowledge transfer. Employing Bayesian theory for the efficient management of task importance weights minimizes the necessity for laborious parameter tuning. By making certain assumptions, the study introduces a deep homoscedastic DA method for enhanced joint task efficacy. Through detailed simulations using a 3D ray tracing dataset, the paper evidences that the integration of environmental semantics and the advanced DA localization techniques markedly elevates the precision of localization in various demanding settings.}

\end{abstract}


\IEEEpeerreviewmaketitle

\section{Introduction}


\IEEEPARstart{L}{ocalization} is becoming an increasingly important part of cellular systems, both because it forms the basis for improved cellular operation (e.g., for planning of handovers), and as a service underlying applications such as intelligent transportation systems. While 5G foresees various ways for ''classical" (e.g., time-of-arrival based) localization, and Beyond 5G (B5G) offers even further expanding opportunities for localization methods \cite{shen2010accuracy, chu2014robust, guo2016accurate,  kanhere2021position, kwon2021joint, conti2021location, zappone2019wireless}, much of the current interest in this field has turned to the ML based localization. 


In this paper, we investigate the demanding localization of users in street canyons in urban areas, which are crucial in intelligent transportation systems, but pose a number of special challenges: (i) they are GPS-denied, thus necessitating a purely cellular solution for localization, (ii) they show high dynamics of the CSI even at fixed locations, due to the movement of vehicles and other objects that can act as scatterers or blockage\cite{alkhateeb2019deepmimo, alrabeiah2020deep}. Consequently, classical localization methods, such as fingerprinting, or trilateration (depending on the existence of a line-of-sight), might fail. As a result, this paper considers ML based localization algorithms that are model-free and can learn novel representations for the data in those scenarios.

\subsection{Related Works}


\emph{Deep Learning based Localization.}
Localization based on deep learning has been shown to improve accuracy and reliability. Consequently, there is a large number of papers on this topic, see, e.g., the surveys \cite{li2019machine, zhu2020indoor, burghal2023comprehensive} and references therein. While many investigations are 
established on received power only, there have also been investigations into the use of the full CSI, i.e., the transfer function or impulse response, possibly observed at multiple antenna elements (or corresponding information in the angular domain). \textcolor{black}{Compared to power-only based localization methods, the fine-grained CSI is more robust to the multi-path fading and temporal dynamics\cite{chu2022sa, molisch2023wireless}. Besides, CSI \cite{wang2015deepfi} and its variants \cite{vieira2017deep, sun2018single} contain detailed information for multi-path characteristics and are thus compatible with a deep structure in the neural network (NN)\cite{li2019machine}}. In deep learning, many types of NNs \cite{vieira2017deep, burghal2023comprehensive} have been employed to represent the fingerprints and offer an end-to-end relationship between the fingerprints and the coordinate(s) of the user(s). For example, \cite{wang2015deepfi} utilized a multiple-layer NN to represent CSI fingerprints, yielding beneficial results in two representative indoor environments. Besides, \cite{sun2018single} considered a convolutional NN and took as its input the angular delay profiles (ADP), whose sparse nature boosts feature extraction. In summary, deep learning technologies pave a new way for localization by providing novel representation for the sampled RF data, building an end-to-end nonlinear relationship between the CSI data and coordinates, and realizing the goal of single-site localization. 

\emph{Deep Domain Adaptation based Transfer Learning for the Enhanced Localization in the Dynamic Environment.}
A special challenge of wireless localization with supervised ML or fingerprinting is that the environment may change between the measurement of the training data and the desired localization. Thus, accommodating time-varying environments is an important goal in algorithm design \cite{mager2015fingerprint,  burghal2023comprehensive}. To address this challenging issue, earlier works proposed utilizing transfer learning (TL) \cite{pan2007adaptive, pan2008transferring}. The related TL based localization algorithms successfully apply to received signal strength (RSS)  data collected over space, devices, and time. Later works employed the CSI data and offered advanced TL methods to improve the robustness of localization against environmental dynamics and other practical constraints \cite{zhang2020transfer, li2021transloc, si2023multi, li2023multi}. However, those novel methods were conducted in an inductive way, in which the network models were trained offline with existing data and retrained with fresh ones. The inductive TL requires online labeled data collection, hindering its practical applications. 

On the other hand, collecting unlabeled data offers more flexibility. DA based TL techniques \cite{zhuang2020comprehensive} enable localizing user(s) accurately and adaptively by utilizing unlabeled RF data, which can be collected easily and cheaply. In the literature, DA methods \cite{chen2020fido, wang2021wifi, chen2022fidora, li2021dafi, chu2022sa, zhou2020adaptive, tian2021wi} were used to improve the scalability and robustness of localization algorithms in a transductive way.
For example, \cite{chen2020fido} formulated the DA framework 
on the basis of variational auto-encoders (VAE). The VAE based DA localization method integrates data argumentation and adaptive knowledge transfer, realizing sub-meter level indoor localization with WIFI CSI data. 
Moreover, our conference paper \cite{chu2022sa} proposed a new adversarial regressive domain adaptation (RDA) method to localize user(s) in a GPS-denied outdoor environment with high dynamics. It adopted an adversarial learning scheme and advanced a nonlinear location regressor to improve the robustness again the environmental changes and the scalability of location space, respectively. Although those methods provide promising localization results, they do not fully take advantage of B5G wireless networks. It is appealing to enhance the localization method with deep learning and B5G wireless networks.

\emph{Semantic Localization.}
\textcolor{black}{The classical model based localization methods are susceptible to the existence of the Line-of-sight (LOS) path and the number of multi-path components (MPCs). For the dynamic scenario, the moving vehicles would temporarily/spatially block the LOS and add/block some non-line of sight (NLOS) path(s). As a result, it is demanding yet essential to reconsider the model based method and exploit a new way, i.e., semantic localization, in this work.} 
Compared to the classical localization setting, semantic localization \cite{schonberger2018semantic, bourdoux20206g, lin2020locater, weber2021semantic} adds the new dimension of location information  -- semantics in the context \cite{bourdoux20206g}. \textcolor{black}{For instance, the integration of visual semantics has been shown to enhance localization performance, as demonstrated in \cite{schonberger2018semantic}, aiding in the removal of ambiguous features. Additionally, \cite{lin2020locater} utilizes semantic subregions (e.g., a conference room within an indoor environment) for location disambiguation.} We note that most semantic localization works have been done for indoor environments, where the semantic context usually refers to the location in a specific room. 

\textcolor{black}{However, localization with the above $static$ semantics is still insufficient in many practical situations, such as GPS-denied and dynamic outdoor environments. The receiver(s) will experience time-varying propagation conditions in a realistic environment caused by moving object(s) between the transmitter and receiver, leading to a dynamic number and amplitude of MPCs. In this work, we take advantage of the CSI and exploit semantic localization for these challenging areas, in which the semantics refers to the main propagation conditions, namely the amount of blockage of the LOS component.} Moreover, the rapid advancement of wireless networks, including B5G cellular networks, enables a more accurate characterization, comprehension, and examination of environmental dynamics \cite{tataria20216g}. Concurrently, the adoption of ML techniques opens up innovative avenues in data modeling \cite{huang2020machine}. 

\subsection{Contribution}

This paper considers both the static and dynamic environmental semantics and aims to make the NN aware of these semantics, striving to realize robust and accurate localization in difficult wireless environments. The key contributions are as follows:
\begin{itemize}
    \item We introduce the propagation condition as a novel indicator for the time-varying environmental semantics and formulate semantic localization as a multi-task learning problem. 
    We note that while for propagation researchers it might seem obvious that the LOS condition impacts the CSI, to our knowledge, this condition has not yet been directly included in any deep learning based end-to-end localization algorithms.   
    \item We propose a multi-task domain adaptation (MDA) based localization scheme, including scenario-adaptive supervised loss functions and novel knowledge alignment designs, which enable the NN to learn good representations for the CSI with both labeled and unlabeled data. The scenario-adaptive knowledge alignment design offers efficient and effective knowledge transfer from labeled data to unlabeled ones, improving the robustness of the localization algorithm against environmental changes. Furthermore, to reduce the time-consuming parameter finetuning in the MDA method, we employ the homoscedastic uncertainty measure in the Bayesian theory to represent the importance weights of each task. Moreover, with some mild assumptions, we attain the log-likelihood of the joint task and solve the problem with the maximum likelihood inference criterion.  
    \item We conduct comprehensive case studies\footnote{To enable reproducibility, the related codes will be released on the project website: \url{https://github.com/Leo-Chu/SemanticLoc}} 
    with a 3D ray tracing dataset, and the experimental results show that semantic localization provides significant improvements compared to the classical one in highly dynamic scenarios. Besides, the proposed localization methods outperform the robustness and accuracy of competing ones in all investigated system configurations. {\textcolor{black}{This suggests their potential for widespread application in outdoor dynamic environments for pedestrian and vehicle positioning}}.      
\end{itemize}

\subsection{Paper organization}
The remainder of this paper is organized as follows. Section \ref{theory_insight} introduces the semantic localization in wireless networks, including the system model and the problem formulation. Then, in Section \ref{section3}, we first elaborate on the proposed MDA localization method and designs of the loss function. Besides, we introduce the improved version to improve the overall robustness. Next, Section \ref{casestudies} extensively compares the proposed localization algorithms with existing ones in different system configurations. Lastly, conclusions are presented in Section \ref{conclusions}. We conclude this section by presenting the symbol notations. 

{\bf Notation} We use bold lower-/upper- case to denote vectors/matrices. The hatted vectors and matrices (${\bf{\hat x}},{\bf{\hat X}}$) refer to their estimates; the notation $\left\|  \cdot  \right\|$ over a vector/matrix means the related ${L}_2$/spectral norm. We use $\mathbb{E}$ to denote the expectation function. The symbols $\Omega$ and $\left|  \Omega  \right|$ represent the data ensemble and its length. The notations $\Omega^S$/${\mathcal{W}}^S$ and $\Omega^T$/${\mathcal{W}}^T$ mean the dataset/parameters in source-domain and target-domain, respectively. Let ${\bf{x}}$ and ${\mathcal{W}}$ be the input and NN parameters (weights and bias), respectively. Then the NN function will be denoted by ${f_{\mathcal{W}}}\left( {\bf{x}} \right)$. The notation $a \propto b$ means $a$ is linearly proportional to $b$. We use $\otimes $ and $\odot$  to denote the Kronecker and Hadamard products.

\section{Semantic Localization in Wireless Networks: Problem Formulation and Theoretical Insight}
\label{theory_insight}

\subsection{Channel Model and Localization Problem}

This work focuses on object positioning, such as pedestrian and self-driving vehicles, in a street canyon, a typical outdoor GPS-denied area relevant for intelligent transportation systems.
We consider a MISO (multiple-input single-output) downlink, such that a user equipment (UE) with a single, possibly omnidirectional, antenna receives  transmissions \footnote{\textcolor{black}{We use the DL and not the UL in order to be in line with the localization philosophy of 3GPP, which in turn is motivated by privacy considerations. Even E911 localization in 3GPP is done in the DL, with multiple BSs transmitting ''positioning reference signals", and the UE performing the actual localization, and feeding back the information if so desired by the UE. For this reason, we are considering a DL situation as well. We note, however, that due to reciprocity of the propagation channels, the same procedure could be applied for the UL as well. Please see, e.g., Dahlman's books on NR \cite{dahlman20225g}, or \cite[Chap 31, 32]{molisch2023wireless}.}} 
from a base station (BS) using \textcolor{black}{orthogonal frequency division multiplexing (OFDM)} modulation. The BS antenna is a uniform planar array with $M_y$ and $M_z$ elements and half-wavelength spacing in horizontal and vertical directions. We assume that the UE can receive the transmit signal from the BS via $P$ MPCs, and each MPC is characterized by complex gain $\alpha_p$, the direction of departure (elevation angle $\theta_p$, and azimuth angle $\phi_p$), and delay $\tau_p$. Then the channel frequency response  (CFR) ${\bf h}[f_k]$ on the $k$-th subcarrier can be denoted by \cite{molisch2023wireless} \vspace{-0.00001cm} 
\begin{equation} 
\label{channel}
{{\bf h}}\left[ {f_k} \right] = \sum\limits_{p = 1}^P {{\alpha _p}{\bf a}\left( {\phi_p ,\theta_p ,{f_k}} \right){e^{ - j2\pi {f_k}{\tau _p}}}} \vspace{-0.00001cm} 
\end{equation}
where ${{\bf a}\left( {{\phi_p},{\theta_p},{f_k}} \right)}$ is the array steering vector of dimension $M \times 1$,  evaluated at the frequency ${f_k}$ and along the direction $\left( {\phi_p ,\theta_p } \right)$. \textcolor{black}{We adopt $M = M_y M_z$ for compact notation.}  Given $K$ available subcarriers and assuming no beam squinting exists (so that ${\bf a}$ becomes independent of $f$), the complete CSI combines all $K$ subcarriers CFR:
${\bf{H}} = \left[ {{\bf{h}}\left[ {{f_1}} \right], \cdots ,{\bf{h}}\left[ {{f_K}} \right]} \right] \in {\mathbb{C}^{M \times N}}.$
For the localization problem 
formulated on the channel model \eqref{channel}, model based algorithms can use the related CSI measurements to infer the distances or the angles between the BS and UE(s) and then apply a simple geometric calculation to get the estimates of coordinates of the UE(s) \cite{zekavat2019handbook}. \textcolor{black}{In our case studies, the moving vehicles will bring in two significant impacts: 1) They may temporally block the LOS path; 2) They may temporally and spatially create/block some new non-LOS path(s). For the channel model in Eq. (1), those impacts are related to the case that the first path may not exist, or we can have a variable number of MPCs ($P$ in Eq. (1)). In this paper, we will use a commercial deterministic channel modeling program, namely \textcolor{black}{Wireless Insite  \cite{Remcom, mededjovic2012wireless}} to generate realistic sets of MPCs and CFRs that form the basis of our simulations. 
 }


\subsection{Semantic Localization}


Localization algorithms are sensitive to environmental dynamics. For example, a slight movement of the UE(s) and/or movement of environmental scatterers will make the phases of the complex amplitudes $arg({\alpha _p})$ vary over space; the $|{\alpha _p}|$, $\left( {\phi_p, \theta_p } \right)$, and $\tau_p$, also change (though more slowly) with the movement of the UE(s). Besides, moving environmental objects, e.g., vehicles, can block MPCs and thus cause significant variations of the absolute amplitudes and may also result in a varying number of MPCs. As a result, it is indispensable to exploit new avenues to robust localization in wireless networks with high dynamics. 

Following the concepts of the modeling theory \cite{weber2021semantic}, one can use a structure $\left\langle {{\bf{P}}, {\bf{d}}} \right\rangle$ to represent the model of the space of semantic localization. The elements of ${\bf{P}}$ refer to the location-related information while the ones of ${\bf{d}}$ provide the spatially meaningful interpretations, i.e., visual semantics \cite{schonberger2018semantic} and subregion index  \cite{lin2020locater}. 

In this work, according to experimental studies \cite{choi2020experimental, chu2022sa}, we use the real-time propagation condition indicators to denote the time-varying geometrical semantics. 
In particular, we distinguish between 1) LOS case, in which the LOS path exists, 2) Dynamic NLOS (DNLOS), where only moving objects such as cars obstruct the LOS, and Static NLOS (SNLOS), where static objects such as houses obstruct the LOS. It is noted that our proposed algorithm imposes no limitations on the number of environmental semantic classes/types. In the case studies, we explore the above-mentioned three different semantic classes. 
\footnote{\textcolor{black}{Note that for the employed ray tracing scenarios in the deepMIMO dataset \cite{alkhateeb2019deepmimo}, the propagation conditions are defined slightly differently, namely 1) LOS; 2) NLOS; 3) Full blockage, which includes no path between the BS and the UE(s). We are however not making use of their classification; we just exclude all data suffering from full blockage since they have all-zero channels.   }}

\subsection{Model based Semantic Localization: Joint Propagation Condition Prediction and Location Estimation}

This section begins by defining the semantic localization problem using a precise statistical CSI model, enabling straightforward location estimation and prediction of related propagation conditions. Various statistics are available to ascertain these propagation conditions \cite{huang2020machine}. \textcolor{black}{In a binary case, such as the identification of LOS and NLOS, one could utilize delay information to signify the propagation conditions.} For example, given $\Gamma$ CSI measurements ${\small {\left\{ {{{{\bf{\tilde H}}}_t}} \right\}_{1, \cdots ,\Gamma}}}$, we are interested in solving a joint estimation problem: \vspace{-0.00001cm}
\begin{equation}
\label{eq2}
\left( {\hat \phi_p ,\hat \theta_p ,\hat \tau_p } \right) = \arg \min \sum\limits_{t = 1}^\Gamma {\left\| {\left( {{{\bf{H}}_t}\left( {\phi_p ,\theta_p ,\tau_p } \right) - {{{\bf{\tilde H}}}_t}} \right)} \right\|_F^2}. 
\vspace{-0.00001cm} 
\end{equation} 
A popular solution to the problem is the space-alternating generalized expectation maximization (SAGE) algorithm  \cite{fessler1994space}. With SAGE, one can estimate 
the related MPCs, i.e., $\hat \phi_p, \hat \theta_p, \hat \tau_p$ in \eqref{eq2}. Then one can easily apply a geometric calculation to get estimates of UE(s)'s coordinates and predefined thresholds (based on the estimated delays) to infer the accompanying propagation conditions \cite{zekavat2019handbook}. 

The literature confirms the advantage of joint estimation, such as integrating both delay and angle information, over approaches that are grounded in a single quantity only approaches \cite{wax1997joint, raleigh1998joint, larsen2009performance, wen2018joint}. For indoor localization, the static semantics can be directly used to enhance the K-nearest neighbor (KNN) algorithm \cite{hernandez2017fuzzy}, which is widely adopted for the final step of location estimation. Similarly, when double-directional channel information  \cite{steinbauer2001double} is given, the combination of angle-of-arrival, angle-of-departure, and delay can improve the performance, as each of these quantities can help to resolve MPCs that cannot be distinguished in one or more of the other dimensions.  However, no off-the-shelf semantics exist in dynamic outdoor scenarios, which motivates us to exploit a new way to determine the appropriate semantics in such an environment.   



\section{Joint Blockage Prediction and Localization: Proposed Methods}
\label{section3}

The joint estimation problem \eqref{eq2} is based on a precise mathematical relationship between the CSI measurements and MPCs. \textcolor{black}{Using estimated MPCs, it is possible to further determine the UE's coordinates (${\left\{ {x\left( t \right),y\left( t \right),z\left( t \right)} \right\}}$) and associated propagation conditions, which are reflected in the delay and its variants. 
Nevertheless, there is a compelling interest in addressing this problem through alternative methods that eliminate the need for an exact signal model, antenna calibration, and the substantial computational demands associated with expectation maximization algorithms.}
\textcolor{black}{Utilizing CSI data (${{{{\bf{\tilde H}}}_t}}$), we employ a NN as a representation tool. This NN is subsequently utilized to concurrently estimate the coordinates of the UE (${I\left( t \right)}$) and identify the associated semantics (${I\left( t \right)}$), such that}
\begin{equation}
\label{eq4}
{{{{\bf{\tilde H}}}_t}} \mathop  \Rightarrow \limits^{NN} \left\{ {\begin{array}{*{20}{c}}
{\left\{ {x\left( t \right),y\left( t \right),z\left( t \right)} \right\}}\\
{I\left( t \right)}
\end{array}} \right.
\end{equation}
\textcolor{black}{To tackle both Coordinates Regression (CR) and Propagation Condition Prediction (PCP)\footnote{In Eq. \eqref{eq4}, drawing from real-scene experiments \cite{choi2020experimental} or the ray-tracing dataset \cite{alkhateeb2019deepmimo}, we concentrate on three types of propagation conditions, depicted as time-varying geometrical semantics.}, the NN must produce outputs for UE's coordinates and the associated PCP, thereby establishing a framework for multi-task learning. Furthermore, we propose innovative designs that leverage these semantics to improve the learning process, thereby enhancing the generalization capabilities of deep learning algorithms. These designs will be detailed in the subsequent sections.} 

\begin{figure}[!t]
\centering
\subfloat[ Training Scheme]{ 
\includegraphics[width=0.475\textwidth]{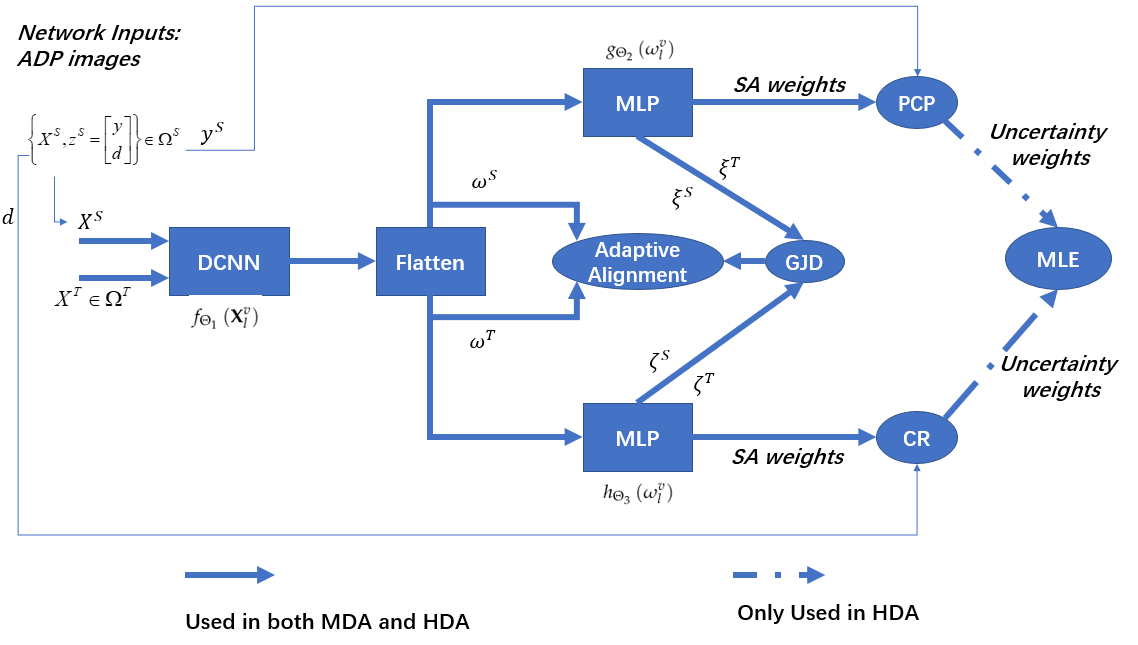}
}\\
\subfloat[ Test Scheme]{
\includegraphics[width=0.475\textwidth]{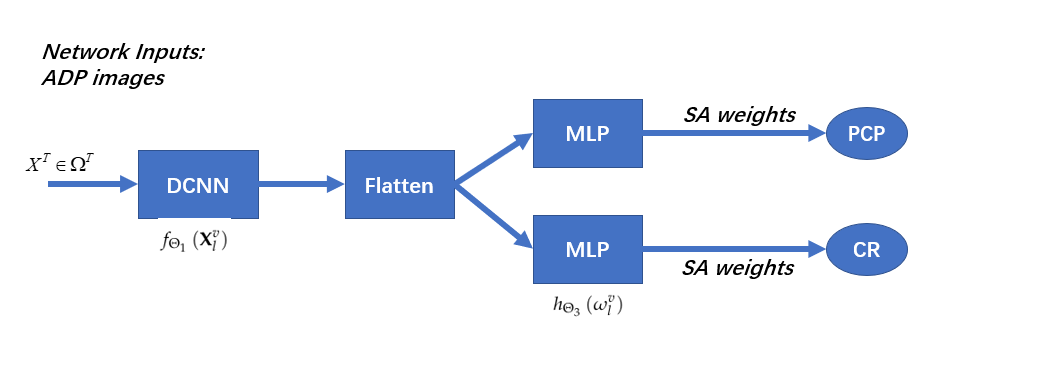}
}
\caption{Schematic overview of the training and testing phases for the proposed methods}
\label{proposedMethod}
\end{figure}


\subsection{On the Multi-task Deep Unsupervised Domain Adaptation based Solution}
\label{MDA_loc}
In this paper, we provide novel localization methods bottomed on the unsupervised DA framework \cite{wilson2020survey}, in which the related deep learning algorithms can learn a unified representation for both labeled and unlabeled data, strengthening the robustness against the effect of high environmental dynamics. Our following designs are motivated by joint estimation  \cite{wax1997joint, raleigh1998joint, wen2018joint}, and multi-task DA theory \cite{tripuraneni2020theory}. \textcolor{black}{In detail, for the semantic localization (including the coordinates estimation and the corresponding semantics identification), we first provide a multi-task deep unsupervised MDA based solution.} 

\textcolor{black}{This section introduces two proposed methods for semantic localization: MDA and Homoscedastic Domain Adaptation (HDA). The MDA method combines multi-task learning and domain adaptation, allowing for regression, semantic identification, and training on both labeled and unlabeled CSI data. This enhances the adaptability of deep learning (DL) based localization across diverse spatiotemporal contexts. Additionally, MDA leverages a semantic indicator to further refine the neural network's training process. The HDA approach, an advanced version of MDA, employs Bayesian theory to model uncertainty in importance weights for supervised tasks, thereby reducing the need for labor-intensive parameter tuning of the weights in the supervised (major) loss functions.}

\subsubsection{Method Overview}

As shown in Fig. \ref{proposedMethod}, we consider the task of unsupervised DA for semantic localization. In this setting, we assume that we have data pairs in the source domain, ${\Omega ^S} = {\left\{ {{\bf{X}}_l^S,{\bf{z}}_l^S} \right\}_{l = 1, \cdots \left| {{\Omega ^S}} \right|,}}$ 
and the target domain,
${\Omega ^T} = {\left\{ {{\bf{X}}_l^T} \right\}_{l = 1, \cdots \left| {{\Omega ^T}} \right|,}}$
where ${\left| {{\Omega ^S}} \right|}$ and ${\left| {{\Omega ^T}} \right|}$ are the number of available data samples in each domain. 
In this paper, \textcolor{black}{to make the notation compact, we define a vector ${\bf{z}} = {\left[ {{{\bf{y}}^T},d} \right]^T}$, where the first three elements of ${\bf{z}}$ are the coordinates of the UE's location (denoted by ${\bf{y}}$) and the last the environmental semantics (the indicator of the propagation condition, denoted by $d$)}. The network input ${\bf{X}}$ denotes the angle-delay domain channel power matrix \cite{sun2018single}, which can be denoted by 
\vspace{-0.000000002cm} 
\begin{equation}
\label{adp2}
{\bf{X}} \buildrel \Delta \over = {\bf{G}} \odot {{\bf{G}}^ * } \in \mathbb{R}^{M \times N}, 
\vspace{-0.000000002cm} 
\end{equation}
where \vspace{-0.000000002cm}  \begin{equation}
\label{adp1}
 {\bf{G}} \buildrel \Delta \over = \frac{1}{{\sqrt {MN} }}\left( {{\bf{V}}_{{M_y}}^H \otimes {\bf{V}}_{{M_z}}^H} \right){\bf{H}}{{\bf{F}}^ * } \in \mathbb{C}^{M \times N},   \vspace{-0.000000002cm} 
\end{equation}
${\left[ {{{\bf{F}}_N}} \right]_{i,k}} = {\textstyle{1 \over {\sqrt N }}}{e^{ - j2\pi {\textstyle{{ik} \over N}}}}$ is the $\left( {i,k} \right)$-th element of the $unitary$ discrete Fourier transform (DFT) matrix ${{\bf{F}}_N} \in {\mathbb{C}^{N \times N}}$, and ${{\bf{V}}_M} \in {\mathbb{C}^{M \times M}}$ denotes a phase-shifted DFT  matrix with the $\left( {i,k} \right)$-th element being ${\left[ {{{\bf{V}}_M}} \right]_{i,k}} = {\textstyle{1 \over {\sqrt M }}}{e^{ - j2\pi {\textstyle{{i\left( {k - M/2} \right)} \over M}}}}$. 
The \textcolor{black}{angular delay profile (ADP)} matrices (${\bf{X}}$) sparsely represent critical features of CSI, making them excellent input candidates that are well suited for the learning process \cite{vieira2017deep, sun2018single}. Please see Sec. \ref{TypesOfInputs} for more types of CSI measurements.

Our goal is to design a hybrid task NN that can reliably and accurately estimate the coordinates of UE(s) and indicate the propagation condition for each CSI measurement (or its variants) in the target domain. To achieve this goal, we present an end-to-end trainable NN, which contains three main network modules: 1) Feature extraction network $f$, parameterized by ${\Theta _1}$; 2) Nonlinear location regression network $g$, parameterized by ${\Theta _2}$; 3) Propagation condition classification network $h$, parameterized by ${\Theta _3}$. The objective function $\mathcal{L}$ for training the proposed network includes four loss terms and is defined as  
\begin{equation}
\label{allLoss}
{\cal L}_{MDA} = \mathop {\min }\limits_{\left\{ {{\Theta _1},{\Theta _2},{\Theta _3}} \right\}} {\lambda _1}{{\cal L}_{{\rm{CR}}}} + {\lambda _2}{{\cal L}_{{\rm{PCP}}}} + {\lambda _3}{{\cal L}_{{\rm{KT}}}} + {\lambda _4}{{\cal L}_{{\rm{WR}}}},  
\end{equation}
where ${\left\{ {{\lambda _k}} \right\}_{k = 1,2,3,4}}$ are the hyperparameters used to control the importance of each loss term. \textcolor{black}{${{\cal L}_{{\rm{CR}}}}$ is used to estimate the coordinates of UE(s) while ${\cal L}_{{\rm {PCP}}}$ identifies the corresponding environmental semantics. These two tasks share the feature extraction module. The scenario-adaptive knowledge alignment loss ${{\cal L}_{{\rm{KT}}}}$ is designed to adaptively optimize the NN with the help of unlabeled data and semantic identification. ${{\cal L}_{{\rm{WR}}}}$ is an auxiliary task to enhance the stability of NN training. }  We will elaborate on each loss function in the following. 

\subsubsection{Scenario-Adaptive Supervised Loss Function for the Joint Task}
\label{SLFJT}
To solve the semantic localization problem shown in Eq. \eqref{eq4}, a natural choice is to use the multiple task learning method, in which we first employ a NN $f_{\Theta _1}$ to perform the feature extraction for the inputs ${{\bf{X}}_l^v}_{v \in \left\{ {S, T}\right\}}$, yielding an output as 
${\left\{\omega_v
\right\}_{v \in \left\{ {S,T} \right\}}} = {f_{{\Theta _1}}}\left( {{\bf{X}}_l^v} \right). $
After that, for the CR task, we use the least square error (L2) loss function to minimize the squared differences between the true coordinates and predicted ones: {\vspace{-0.000000002cm} 
\begin{align}
\label{CR}
{{\cal L}_{\rm CR}} &=  {\mathbb{E}_{\left\{ {{{\bf{X}}^S},{{\bf{y}}^S}} \right\} \in {\Omega ^S}}}{L_2}\left( {{\Theta _1},{\Theta _2}} \right) \nonumber \\
 &=   {\frac{1}{{\left| {{\Omega ^S}} \right|}}\sum\nolimits_l {\left\| {\zeta_l ^v - {\bf{y}}_l^S} \right\|_2^2} }. \vspace{-0.000000002cm}  
\end{align}}

Furthermore, we calculate the modulated cross-entropy (CE) loss (also known as the focal loss) \cite{lin2017focal} between the PCP task predictions and corresponding ground truth labels $d_l$ as  { \vspace{-0.000000002cm} 
\begin{align}
\label{PCP}
{{\cal L}_{\rm PCP}} &=   {\mathbb{E}_{\left\{ {{{\bf{X}}^S},{d^S}} \right\} \in {\Omega ^S}}}{L_{\rm CE}}\left( {{\Theta _1},{\Theta _3}} \right) \nonumber \\
 &=  {\frac{1}{{\left| {{\Omega ^S}} \right|}}\sum\nolimits_l {\rm CE\left( -{{\left( {1 - \xi_l ^v} \right)^\gamma }\log \xi_l ^v,d_l^S} \right)}}.
\vspace{-0.000000002cm}  \end{align}}
The loss functions in \eqref{CR} and \eqref{PCP} are used for training ${f_{{\Theta _1}}}$ and the two specific networks ${g_{{\Theta _2}}}$ and ${h_{{\Theta _3}}}$ with labeled data. For brevity of notation, in \eqref{CR} and \eqref{PCP}, we have $\zeta_l ^v{\rm{ = }}{g_{{\Theta _2}}}\left( {\omega_l ^v} \right), \ \xi_l ^v = {h_{{\Theta _3}}}\left( {\omega_l ^v} \right), v \in \left\{S\right\}$. ${\left( {1 - \xi ^v} \right)^\gamma }$ is the modulating factor \cite{lin2017focal} with the non-negative number $\gamma$ being its $temperature$, which is also known as the focusing parameter. The loss ${{\cal L}_ {\rm PCP}}$ is equivalent to the vanilla CE when $\gamma=0$. The loss function shown in \eqref{PCP} is meaningful in practice. Consider a case with semantic label 1 (NLOS CSI samples), and $\xi_l ^v$ small, then the modulating factor is near 1, which means the loss ${{\cal L}_{\rm PCP}}$ is unaffected. On the other hand, when we have a case with the semantic label being 0, during the training process, the modulating factor goes to zero ($\xi _l^v \to 1$), and the loss for LOS CSI samples is down-weighted. As a result, with the modulated CE loss   \eqref{PCP}, the network optimization will be conducted in an adaptive way, granting more weights for NLOS (hard to learn) samples. Assigning different importance weights for examples with different semantic labels is one of the 
key designs for semantic localization.           

\subsubsection{Scenario-Adaptive Knowledge Alignment Loss Function}

Based on the framework of deep unsupervised DA, our goal is to align the knowledge learned from labeled data to unlabeled ones. The popular implementations minimize the distance between learning representations from the source and target domains. For example, the distance measure \cite{wilson2020survey} can be Maximum Mean Discrepancy, Kullback–Leibler (KL) divergence, and their variants. In our case, for data in the source and target domains, we have one local representation ${\left\{ {{\omega ^v}} \right\}_{v \in \left\{ {S, T} \right\}}}$ and two global representations: ${g_{{\Theta _2}}}\left( {\omega^v} \right)$ and ${h_{{\Theta _3}}}\left( {\omega^v} \right)$. 

However, for semantic localization, fully matching the whole distributions of source and target CSI measurements to each other at the global level may not be effective, as domains could have scene layouts, i.e., different moving vehicle positions.       
On the other hand, adaptive alignment yields superior performance derived from recent advances in DA theory and related applications \cite{long2018conditional, saito2019strong}. For example, the strong-weak distribution alignment \cite{saito2019strong} and adversarial matching strategies \cite{long2018conditional} significantly improve the DA algorithms' performance in objection detection and image recognition tasks. This motivates us to propose a scenario adaptive knowledge alignment for semantic localization. We first employ the symmetrical Kullback–Leibler (SKL) divergence to align the local features, which can be denoted by { \vspace{-0.000000002cm} 
\begin{equation}
\label{loc_align}
{L_{\rm loc}} =   {\mathbb{E}_{{\bf{X}} \in \left\{ {{\Omega ^S} \cup {\Omega ^T}} \right\}}}SKL\left( {{\omega ^S},{\omega ^T}} \right), \vspace{-0.000000002cm}
\end{equation}}
where \[SKL\left( {{\omega ^S},{\omega ^T}} \right) = {D_{KL}}\left( {{\omega ^S}||{\omega ^T}} \right) + {D_{KL}}\left( {{\omega ^T}||{\omega ^S}} \right), \] \[{\omega ^v} = \frac{{P_v^{'}}}{{\sum {P_v^{'}} }},P_v^{'} = \frac{1}{{\left| {{\Omega ^v}} \right|}}\sum\nolimits_l {\omega _l^v} ,v \in \left\{ {S,T} \right\}, \]
and ${D_{KL}}\left( {P||Q} \right) = \sum\nolimits_i {P\left( i \right)} \ln \frac{{P\left( i \right)}}{{Q\left( i \right)}}.$

For the global feature alignment, we have 
\begin{equation}
\small
\label{globe_align}
{L_{\rm global}} =  {\mathbb{E}_{{\bf{X}} \in \left\{ {{\Omega ^S} \cup {\Omega ^T}} \right\}}}SKL\left( {{\Phi _ \otimes }\left( {{\xi ^S},{\zeta ^S}} \right),{\Phi _ \otimes }\left( {{\xi ^T},{\zeta ^T}} \right)} \right), 
\end{equation} 
where $\otimes$ denotes the Kronecker product, and we use the modulated multilinear condition map \cite{song2009hilbert} to get the joint distribution of global representations in the CR and PCP task:  { \vspace{-0.000000002cm} 
\begin{equation}
\label{mcp}
{\Phi _ \otimes }\left( {\xi ^v,\zeta ^v} \right) = -{\left(1 - { \xi ^v} \right)^\gamma }\log (\xi^v)  \otimes \zeta ^v,v \in \left\{ {S,T} \right\}. \vspace{-0.000000002cm} 
\end{equation} }
The multilinear condition map \eqref{mcp} captures the global multi-modal conformation behind complex data distributions \cite{long2018conditional}. 
Besides, it is designed to put more weight on the NLOS samples than the LOS ones during the training and to dominate the distribution of global representation in the PCP task with LOS samples, making this novel map scenario adaptive. Since the SKL is a convex function in the domain of probability distributions, the alignment based on the SKL is also scenario adaptive. 

Putting everything together, we get the scenario adaptive knowledge alignment loss function as  
\vspace{-0.000000002cm} 
\begin{equation}
\label{KTLoss}
{{\cal L}_{{\rm{KT}}}} = {L_{\rm global}} + {L_{\rm loc}}. 
\vspace{-0.000000002cm} 
\end{equation}



Lastly, let ${\Theta  = \left\{ {{\Theta _1},{\Theta _2},{\Theta _3}} \right\}}$, we use an auxiliary weight regularization (WR),  
\begin{equation}
{\cal L_{{\rm{WR}}}} = {\mathbb{E}_{{\bf{X}} \in \left\{ {{\Omega ^S} \cup {\Omega ^T}} \right\}}}{\textstyle{1 \over 2}}\left\| \Theta  \right\|_2^2 
\end{equation}
to avoid over-fitting and enhance the robustness of deep NN models\cite{goodfellow2016regularization}.

\subsection{Semantic Localization With Homoscedastic Domain Adaptation}

\subsubsection{Motivations}

\begin{figure}[!t]
\centering
\subfloat[\label{figMPCs1} System I]{ 
\includegraphics[width=0.475\textwidth]{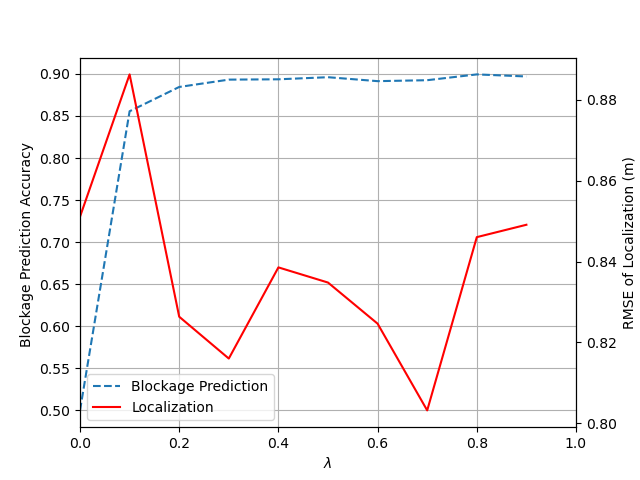}
}\\
\subfloat[\label{figMPCs2} System II]{
\includegraphics[width=0.475\textwidth]{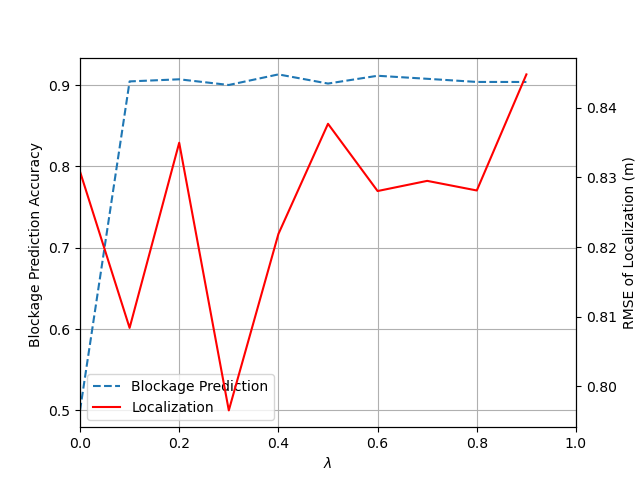}
}
\caption{The performance variation of the proposed semantic localization algorithm (MDA) over different values of $\lambda$. Here, we assume $\lambda_1 + \lambda_2 = 1$ and leave out the subscript for simplicity. Case I and Case II denote a communication system with the related bandwidth being 100M and 500M, respectively.}
\label{lambda}
\end{figure}

\textcolor{black}{This subsection presents a method for the automatic fine-tuning of the weights ${\lambda _1}$ and ${\lambda _2}$ in supervised (dominant) loss functions. This method is motivated by the observation that the effectiveness of the localization algorithm is heavily influenced by the balance between the weights assigned to each dominant (supervised) task loss (${{\cal L}_{{\rm{CR}}}}, \ {{\cal L}_{{\rm {PCP}}}}$).} As demonstrated in Fig. \ref{lambda}, adjustment of these weights is necessary to optimize performance across different communication system configurations, which - if it has to be done manually - adds to the computational load. \textcolor{black}{Consequently, our method aims to learn how to simultaneously balance the regression and classification tasks under varying conditions. This approach minimizes the need for manual weight adjustments that are typically required in the MDA method.}

\textcolor{black}{The literature, including works \cite{ byrd2019effect, garg2020functional, tripuraneni2020theory}, emphasizes that a NN's generalization ability is significantly linked to the amount of labeled and unlabeled data it is trained on. Generalization, measured by uniform stability, reveals that the generalization capacity of NNs is primarily influenced by the quantity of labeled data, with the addition of unlabeled data further enhancing this generalization. Consequently, our proposed HDA method concentrates on automatically adjusting the first two weights, focusing on optimizing these crucial parameters rather than adjusting all weights.}

{\textcolor{black}{{\em Remark:} In our case studies, we explored several dynamic methods for optimizing weights, including fixed weights, linear updating, and scheduled learning strategies, drawing inspiration from different regularization strategies for deep learning \cite{goodfellow2016regularization}. Our findings reveal that an exponential updating method marginally outperforms the other approaches. Furthermore, our approach, alongside the strategy mentioned in reference \cite{long2018conditional}, aligns with the domain adaptation subset of transfer learning. This alignment suggests that employing a similar strategy for optimizing weights related to unlabeled data ($\lambda_3$) could be effective, as evidenced by its success across various datasets. Additionally, the regularization technique serves as an auxiliary tool for training NNs, chiefly designed to mitigate overfitting and enhance the robustness of deep NN models, as outlined in \cite{goodfellow2016regularization}. Despite its recognition and utility in existing studies, the implementation of ${\cal L_{WR}}$ yields only a slight improvement in performance on test datasets. Consequently, we do not propose an automatic tuning solution for $\lambda_4$.}} 

\subsubsection{Maximum Likelihood Estimation in the Joint Task}

To solve this challenging issue, we employ the related Bayesian learning theory \cite{kendall2017uncertainties} to model the weights (also known as the relative confidence) in the CR and PCP tasks with some uncertainty measures, which will be elaborated in the following. Since these weights are task-independent, the corresponding uncertainty measures and the knowledge transfer technology will be referred to as Homoscedastic uncertainty \cite{kendall2017uncertainties} and HDA. Our goal is to obtain the likelihood function of the joint task, as shown in Section \ref{SLFJT}. We make some mild assumptions about the network outputs to tackle this difficult issue. 

\begin{assumption}
\label{ass1}
Let ${\bf x}$ and ${\bf y}$ be the input data and corresponding labels of a NN $g$, parameterized by ${\bf{\mathcal{W}_1}}$. Denoting the random output of the network model as ${g_{\bf{\mathcal{W}_1}}}\left( {\bf{x}} \right)$, then the model likelihood is given by $p\left( {{\bf{y}}|{g_{\bf{\mathcal{W}_1}}}\left( {\bf{x}} \right)} \right)$. For regression tasks, one can define the likelihood as a Gaussian with the mean given by the model output \cite{kendall2017uncertainties}
\vspace{-0.000000002cm} 
\begin{equation}
p\left( {{\bf{y}}|{g_{\bf{\mathcal{W}_1}}}\left( {\bf{x}} \right)} \right) = \mathcal{N}\left( {{g_{\bf{\mathcal{W}_1}}}\left( {\bf{x}} \right),{\sigma_1 ^2}} \right),  \vspace{-0.000000002cm}
\end{equation}  
with $\sigma_1 ^2$ being an observation noise scalar. 
\end{assumption}

\begin{assumption}
\label{ass2}
Let ${\bf x}$  be the input of a multiple-task NN with $K$ outputs, denoted by ${\left\{ {{{\bf{y}}_k}} \right\}_{k = 1, \cdots ,K}}$. These outputs are independent events.
\end{assumption}


Following the analysis in Section \ref{MDA_loc}, the learnable network parameter ensembles in the PCP and CR tasks are denoted by ${\bf{\mathcal{W}_1}}=\left\{ {{\Theta _1},{\Theta _2}} \right\}$, ${\bf{\mathcal{W}_2}}=\left\{ {{\Theta _1},{\Theta _3}} \right\}$, respectively. The unknown variable $\sigma_1 ^2$ will be learned during the optimization. With these mild assumptions, we can obtain the following result.
\begin{theorem}
\label{theorem1}
Let $\Theta$ be the learnable parameters ensemble in the investigated NN. The network model's outputs constitute a continuous output ${{\bf{y}}_1}$ (CR task) and a discrete one ${{\bf{y}}_2}$ (PCP task with standard CE loss). The related log-likelihood of the joint task can be given by:
\begin{equation}
\label{MLJK}
\small
\begin{array}{l}
\log p\left( {{{\bf{y}}_1},{{\bf{y}}_2}|{f_{\Theta }}\left( {\bf{x}} \right)} \right) \propto  - \frac{1}{{2\sigma _1^2}}{{\cal L}_{\rm{1}}} - \frac{1}{{\sigma _2^2}}{{\cal L}_{\rm{2}}} - \log {\sigma _1} - \log \sigma _2^2,
\end{array}
\end{equation}
where ${{L}_{\rm{1}}} = {\left\| {{{\bf{y}}_1} - {g_{\bf{{\mathcal{W}_1}}}}\left( {\bf{x}} \right)} \right\|^2}$, and ${{L}_{\rm{2}}} =  -\log {\rm{softmax}}\left({{h_{\bf{{\mathcal{W}_2}}}}\left( {\bf{x}} \right)} \right)$, ${\left\{ {{\sigma _k}} \right\}_{k = 1,2}}$ are some learnable probability rescaling factors.
\end{theorem}

{\bf Proof:} Since we have the assumption for the output of the CR network branch, we start proving the result in Theorem \ref{theorem1} by focusing on the likelihood function of the PCP network branch, which can be denoted by\vspace{-0.000000002cm}
\begin{equation}
\label{app11}
p\left( {{{\bf{y}}_{\bf{2}}}|\frac{1}{{\sigma _2^2}}{h_{{{\cal W}_{\bf{2}}}}}\left( {\bf{x}} \right)} \right) = {\rm{softmax}}\left( {\frac{1}{{\sigma _2^2}}{h_{{{\cal W}_{\bf{2}}}}}\left( {\bf{x}} \right)} \right), \vspace{-0.000000002cm}
\end{equation}
where the related network parameter ensemble ${\bf{\mathcal{W}_2}}$ was given in Section \ref{SLFJT} (${\bf{\mathcal{W}_2}}=\left\{ {{\Theta _1},{\Theta _3}} \right\}$), and ${{\sigma _2^2}}$ is an auxiliary variable to denote the probability (temperature) rescaling level \cite{guo2017calibration}. Eq.\eqref{app11} boils down to the vanilla softmax function when ${{\sigma _2^2}}=1$. Suppose that the output of the $\rm{softmax}$ function gives the $i$-th class, then the corresponding log-likelihood yields \vspace{-0.00001cm}
\begin{equation}
\label{app12}
\log p\left( {{{\bf{y}}_2} = i|{h_{{{\cal W}_{\bf{2}}}}}\left( {\bf{x}} \right)} \right) = \frac{1}{{{\sigma ^2}}}{{\bf{u}}_i} - \log \sum\nolimits_j {\frac{1}{{{\sigma ^2}}}{{\bf{u}}_{ij}}},  
\end{equation}
With some manipulation, we can simplify Eq.\eqref{app12} as
\begin{equation}
\label{app121}
\begin{array}{*{20}{l}}
{\log p\left( {{{\bf{y}}_2} = i|\frac{1}{{\sigma _2^2}}{\bf{u}}} \right)}\\
{ = \frac{1}{{\sigma _2^2}}\left( {{{\bf{u}}_i} - \sigma _2^2\log \sum\nolimits_j {\exp \left( {\frac{1}{{\sigma _2^2}}{{\bf{u}}_{ij}}} \right)} } \right)}\\
{ = \frac{1}{{\sigma _2^2}}\left( {{{\bf{u}}_i} - \log \left[ {\sum\nolimits_j {\exp \left( {{{\bf{u}}_{ij}}} \right)} } \right]} \right)}\\
{ + \left( {\frac{1}{{\sigma _2^2}}\log \left( {\sum\nolimits_j {\exp \left( {{{\bf{u}}_{ij}}} \right)} } \right) - \log \sum\nolimits_j {\exp \left( {\frac{1}{{\sigma _2^2}}{{\bf{u}}_{ij}}} \right)} } \right)}\\
{ = p\left( {{{\bf{y}}_{\bf{2}}}|{{\bf{u}}_i}\left( {\bf{x}} \right)} \right) + \log \frac{{{{\left( {\sum\nolimits_j {\exp \left( {{{\bf{u}}_{ij}}} \right)} } \right)}^{\frac{1}{{\sigma _2^2}}}}}}{{\sum\nolimits_j {\exp \left( {\frac{1}{{\sigma _2^2}}{{\bf{u}}_{ij}}} \right)} }}}\\
{\mathop  \approx \limits^{{assump.}} \log {\rm{softmax}}\left( {{{\bf{u}}_i}} \right) - \log \sigma _2^2}
\end{array}
\end{equation}
In Eq.\eqref{app121}, $assump.$ denotes the case we have ${\left( {\sum\nolimits_j {\exp \left( {{{\bf{u}}_{ij}}} \right)} } \right)^{\frac{1}{{\sigma _2^2}}}} \approx \frac{1}{{\sigma _2^2}}\sum\nolimits_j {\exp \left( {\frac{1}{{\sigma _2^2}}{{\bf{u}}_{ij}}} \right)} $, which becomes an equality when $\sigma _2^2 \to 1$.
With Assumption \ref{ass2}, we can have 
\begin{equation}
\label{app131}
\begin{array}{l}
\log p\left( {{{\bf{y}}_1}|{f_{{{\bf{\mathcal W}}_1}}}\left( {\bf{x}} \right)} \right)\\
 = \log \left( {\frac{1}{{{\sigma _1}\sqrt {2\pi } }}\exp \left( { - \frac{1}{2}\frac{{{{\left\| {{{\bf{y}}_1} - {f_{\bf{\mathcal W}_1}}\left( {\bf{x}} \right)} \right\|}^2}}}{{\sigma _1^2}}} \right)} \right)\\
 \propto  - \frac{1}{{2\sigma _1^2}}{\left\| {{{\bf{y}}_1} - {f_{\bf{\mathcal W}_1}}\left( {\bf{x}} \right)} \right\|^2} - \log {\sigma _1}
\end{array}
\end{equation}

Lastly, together with Eq.\eqref{app121}-\eqref{app131} and Assumption \ref{ass2}, we are ready to get the log-likelihood of the output of the NN model (${q_{\bf{W}}}\left( {\bf{x}} \right)$):
\begin{equation}
\label{app13}
\begin{array}{l}
\log p\left( {{{\bf{y}}_1},{{\bf{y}}_2} = i|{q_{\bf{W}}}\left( {\bf{x}} \right)} \right)\\
 = \log p\left( {{{\bf{y}}_1}|{g_{{W_{\bf{1}}}}}\left( {\bf{x}} \right)} \right)p\left( {{{\bf{y}}_2} = c|{h_{{W_{\bf{2}}}}}\left( {\bf{x}} \right)} \right)\\
 = \log p\left( {{{\bf{y}}_1}|{g_{{W_{\bf{1}}}}}\left( {\bf{x}} \right)} \right)p\left( {{{\bf{y}}_2} = c|{h_{{W_{\bf{2}}}}}\left( {\bf{x}} \right)} \right)\\
 \propto  - \frac{1}{{2\sigma _1^2}}{\left\| {{{\bf{y}}_1} - {g_{{W_{\bf{1}}}}}\left( {\bf{x}} \right)} \right\|^2} - \log {\sigma _1} - \log \sigma _2^2\\
 + \frac{1}{{\sigma _2^2}}\log {\rm{softmax}}\left( {\bf{u}} \right)
\end{array},
\end{equation}
which completes the proof of Theorem \ref{theorem1}.
\begin{corollary}
\label{corollary1}
Let ${\bf{u}} = {{h_{{{\cal W}_{\bf{2}}}}}\left( {\bf{x}} \right)} $, for a joint task NN with ${{\bf{y}}_1}$ and ${{\bf{y}}_2}$ being its outputs, applying modulated CE loss in the PCP task yields a log-likelihood as:
\begin{equation}
\label{MLJK2}
\begin{array}{l}
\log p\left( {{{\bf{y}}_1},{{\bf{y}}_2} = c|{f_\Theta }\left( {\bf{x}} \right)} \right)\\
 \propto  - \frac{1}{{2\sigma _1^2}}{\left\| {{{\bf{y}}_1} - {f_{{{\cal W}_{\bf{1}}}}}\left( {\bf{x}} \right)} \right\|^2} - \log {\sigma _1} - \log \sigma _2^2\\
 + {\left( {1 - \frac{1}{{\sigma _2^2}}{{{\rm{softmax}} \left( {\bf{u}} \right)}^{\frac{1}{{\sigma _2^2}}}}} \right)^\gamma } {\log \left( {{\rm{softmax}}\left( {\bf{u}} \right)} \right)}
\end{array}.
\end{equation}
\end{corollary}

{\bf Proof:} For the modulated CE loss (Eq. \eqref{PCP}), we are given a weighted conditional distribution function, 
\begin{equation}
\label{app21}
- {\left( {1 - p\left( {{{\bf{y}}_2} = c|\frac{1}{{\sigma _2^2}}{\bf{u}}} \right)} \right)^\gamma }\log p\left( {{{\bf{y}}_2} = c|\frac{1}{{\sigma _2^2}}{\bf{u}}} \right),
\end{equation}
where ${\bf{u}} = {{h_{{{\cal W}_{\bf{2}}}}}\left( {\bf{x}} \right)}$. Our goal here is to simplify the likelihood function $p\left( {{{\bf{y}}_2} = c|\frac{1}{{\sigma _2^2}}{\bf{u}}} \right)$ and related log-likelihood function.
Similar to Eq.\eqref{app11}, we have
\begin{equation}
\label{app23}
\begin{array}{l}
p\left( {{{\bf{y}}_{\bf{2}}} = i|\frac{1}{{\sigma _2^2}}{\bf{u}}} \right) = {\rm{softmax}}\left( {\frac{1}{{\sigma _2^2}}{\bf{u}}} \right)\\
 = \frac{{{{\left( {\exp \left( {{{\bf{u}}_i}} \right)} \right)}^{\frac{1}{{\sigma _2^2}}}}}}{{\sum\nolimits_j {\exp \left( {\frac{1}{{\sigma _2^2}}{{\bf{u}}_{ij}}} \right)} }} \approx \frac{{{{\left( {\exp \left( {{{\bf{u}}_i}} \right)} \right)}^{\frac{1}{{\sigma _2^2}}}}}}{{\sigma _2^2{{\left( {\sum\nolimits_j {\exp \left( {{{\bf{u}}_{ij}}} \right)} } \right)}^{\frac{1}{{\sigma _2^2}}}}}} = \frac{1}{{\sigma _2^2}}{\left( {{\rm{softmax}}\left( {\bf{u}} \right)} \right)^{\frac{1}{{\sigma _2^2}}}}
\end{array},
\end{equation}

Following the analysis shown in Eq.\eqref{app121}, we have 
\begin{equation}
\label{app22}
\begin{array}{*{20}{l}}
\begin{array}{l}
\log p\left( {{{\bf{y}}_2} = i|\frac{1}{{\sigma _2^2}}{\bf{u}}} \right)\\
 \approx \frac{1}{{\sigma _2^2}}\log p\left( {{{\bf{y}}_2} = c|{\bf{u}}} \right) - \log \sigma _2^2
\end{array}\\
{ = \log {\rm{softmax}}\left( {\bf{u}} \right) - \log \sigma _2^2}
\end{array}.
\end{equation}
Plugging the results shown in Eq.\eqref{app121} and Eq.\eqref{app23} into Eq.\eqref{app21} gives
\begin{equation}
\label{app24}
{\left( {1 - \frac{1}{{\sigma _2^2}}{{\left( {{\rm{softmax}}\left( {\bf{u}} \right)} \right)}^{\frac{1}{{\sigma _2^2}}}}} \right)^\gamma }\left( {\log {\rm{softmax}}\left( {\bf{u}} \right) - \log \sigma _2^2} \right).
\end{equation}
Similar to the analysis in Eq.\eqref{app13}, we arrive to the result in Eq.\eqref{MLJK2} with Eq.\eqref{app131},  Eq.\eqref{app24}, and Assumption \ref{ass2}.

We have two remarks hinged on Theorem \ref{theorem1} and Corollary \ref{corollary1}.
\begin{remark}
The log-likelihood of the joint task (\eqref{MLJK} or \eqref{MLJK2}) is a linear combination of the loss functions ${\cal L}_{\rm PCP}$ and ${\cal L}_{CR}$.
\end{remark}
\begin{remark}
With the mean squared error (MSE) criterion, the optimization is implemented by minimizing the MSE of the targets and related estimates. Similarly, in the maximum likelihood inference, one can maximize the log-likelihood of the model. In the literature, these two optimization schemes are effective for a similar optimization target with different principles \cite{jordan1999introduction, chu2019efficient}.
\end{remark}

In this work, we employ the maximum likelihood principle to perform optimization for a joint task NN while measuring the uncertainty of each task with two learnable probability rescaling factors. The factor ${\sigma _1}$ in the CR task captures how much noise we have in the outputs, while ${\sigma _2}$ in the PCP task describes the confidence of the predictions \cite{guo2017calibration}. These factors can be learned \cite{guo2017calibration} or derived using statistical methods \cite{chu2019eigen}. Minimization of ${\sigma _1}$ and ${\sigma _2}$ are related to surpassing the noise and alleviating the negative effect of overconfident predictions.   

\subsubsection{Overall Loss}

With Theorem \ref{theorem1}, we will cooperatively optimize the NN by minimizing the negative log-likelihood of the joint task (Eq. \eqref{MLJK2}). Besides, the relative confidence for the CR and PCP tasks are determined by two learnable parameters ${{\sigma _1}}$ and ${{\sigma _2}}$ and will be optimized during the whole algorithm optimization. We use the same knowledge alignment loss function and weight regularization shown in Section. \ref{MDA_loc}. Based on the above analysis, we finally get the overall loss for the proposed HDA localization method, denoted by
\begin{equation}
\label{HDA_allLoss}
\small
{\mathcal{L}_{HDA}} = \mathop {\min }\limits_{\left\{ {\Theta ,{\sigma _1},{\sigma _2}} \right\}}  - \log p\left( {{{\bf{y}}_1},{{\bf{y}}_2} = c|{f_\Theta }\left( {\bf{x}} \right)} \right) + {\lambda _3}{\mathcal{L}_{{\rm{KT}}}} + {\lambda _4}{\mathcal{L}_{{\rm{WR}}}}.
\end{equation}

\begin{table*}[t]
\centering
\caption{\textcolor{black}{Notes on the adopted system parameters in O2 Dynamic Scenario in deepMIMO dataset\cite{alkhateeb2019deepmimo}.}}
\begin{tabular}{|l|l|l|l|}
\hline
Parameters             &  Default  & Section with Default Setting   & Change notes in Section                                                                      \\ \hline
Carrier Frequency     & 3.5 GHz           &  Section \ref{Semantic1} and \ref{convergence} &                                                                                  \\ \hline
Grid size              & 0.8m             &  Section \ref{Semantic1} and \ref{convergence}   &                                                                                 \\ \hline
BS Location            & (3,10,6)         &  Section \ref{Semantic1} and \ref{convergence}    &                                                                                \\ \hline
Subcarriers  & 64               &      Section \ref{Semantic1} and \ref{convergence}      &                                                                          \\ \hline
Antenna                & $8\times8$ Planar Array &        Section \ref{Semantic1} and \ref{convergence}    &       Section \ref{TypesOfInputs}                       \\ \hline
MPCs & 25               & Section \ref{Semantic1} and \ref{convergence} & Section \ref{expMPCs}  \\ \hline
Bandwidth              & 100 MHz           &                    Section \ref{Semantic1} and \ref{convergence}    &           Section \ref{expBand}             \\ \hline
Input type             & ADP              &  Section \ref{Semantic1} and \ref{convergence} &  Section \ref{TypesOfInputs}                                     \\ \hline
Normalization          & Antenna-wise     & Section \ref{Semantic1} and \ref{convergence}                             &      Section \ref{expNorm}            \\ \hline
\end{tabular}
\label{systemPara}
\end{table*}

The MDA method is designed for the novel idea of semantic localization, which adds wireless semantics analysis to assist the localization in a challenging environment (with high dynamics and weak GPS signals). We first solve related semantic localization with the MDA method, which integrates two independent tasks with the unsupervised DA framework. Furthermore, to avoid using a naive weighted sum of supervised (dominant) losses or manually tuning the related weights, we propose the HDA method, which combines multiple loss functions with homoscedastic task uncertainty. Comparing the design shown in \eqref{allLoss}, \eqref{HDA_allLoss} only adds two more learnable parameters. However, it significantly reduces the computational complexity of the MDA method in wireless networks with several changeable system parameters, such as time-varying MPCs, bandwidths, and carrying frequencies. 


\section{Case Studies}
\label{casestudies}

\textcolor{black}{
This section delves into case studies where we first introduce the implementation of neural networks and the corresponding datasets. Following this introduction, we evaluate the efficacy of several deep learning models across diverse wireless system specifications that impact the datasets. Lastly, we assess the efficiency of deep learning methods by presenting convergence and complexity analysis.}

\subsection{The Neural Network Implementation, Dataset, and Evaluation Protocols}

\begin{table*}[ht]
\centering
	\small
	\begin{tabular}{|l|ll|l|l|l|l|}
		\hline
		Learning Task                                        & \multicolumn{2}{l|}{\begin{tabular}[c]{@{}l@{}}Main     Task    \\ Weight\end{tabular}} & \begin{tabular}[c]{@{}l@{}} KT \\    Weight\end{tabular} & \begin{tabular}[c]{@{}l@{}}Localization\\    (RMSE)\end{tabular} & \begin{tabular}[c]{@{}l@{}} Propagation Condition  \\ Prediction(PCP) (Accuracy)\end{tabular} & \begin{tabular}[c]{@{}l@{}}Performance Gains \\ (Localization/PCP)   \end{tabular} \\ \hline
		\multirow{2}{*} {\begin{tabular}[c]{@{}l@{}}Only Location \\ Regression \end{tabular}}     & \multicolumn{1}{l|}{1}                                   & 0                                   &  \ding{55}                                                                                   & 1.0567                                                                   & -                                                                                        & -                                                                 \\ \cline{2-7} 
		& \multicolumn{1}{l|}{1}                                   & 0                                   & \checkmark                                                                                  & 0.9893                                                                   & -                                                                                        & $\bf{\%6.813}$                                                                \\ \hline
		\multirow{2}{*}{\begin{tabular}[c]{@{}l@{}}Only Blockage \\ Indication \end{tabular}} & \multicolumn{1}{l|}{0}                                   & 1                                   & \ding{55}                                                                                 & -                                                                   & 0.8772                                                                                        & -                                                                 \\ \cline{2-7} 
		& \multicolumn{1}{l|}{0}                                   & 1                                   & \checkmark                                                                                    & -                                                                   & 0.9148                                                                                       & $\bf{\%4.286}$                                                                \\ \hline
		Unweighted                                  & \multicolumn{1}{l|}{0.5}                                 & 0.5                                 & \checkmark                                                                                  & 0.8802                                                                   & 0.9311                                                                                        & $\bf{\%20.05}$ /$\bf{\%6.145}$                                                                 \\ \hline
		Approx. Opt. (MDA)                   & \multicolumn{1}{l|}{0.7}                                 & 0.3                                 & \checkmark                                                                                  & 0.8299                                                                   & 0.9417                                                                                        & $\bf{\%27.33}$ /$\bf{\%7.353}$                                                                  \\ \hline
		\begin{tabular}[c]{@{}l@{}}Uncertainty \\ weighting (HDA) \end{tabular}                        & \multicolumn{1}{l|}{\checkmark}                                   & \checkmark                                   & \checkmark                                                                                  & 0.8165                                                                   &  0.9359                                                                                       &  $\bf{\%29.42}$ /$\bf{\%6.692}$                                                                \\ \hline
	\end{tabular}
\caption{\label{tabx1} Quantitative experiments for each loss function in the proposed method.  \checkmark/\ding{55} means "with"/"without";  ‘-’ means None. BP/KT denotes Blockage Prediction/Knowledge Transfer.}
\end{table*}

\subsubsection{Implementation Details}

We implement our network model using Pytorch and train the network using Google Colab with up to 52GB graphics processing unit (GPU) memory. 
To enable fair comparison, we use a similar network structure as in \cite{vieira2017deep} to perform feature extraction ($\Theta_1$) for all compared methods. The structure of $\Theta_1$ consists of four convolutional layers, each followed by a 2D batch normalization functional, max pooling, and a ReLU activation. For the non-linear location regression network $\Theta_2$ and the propagation condition classification $\Theta_3$, we use the Multilayer perceptron (MLP) structure. We employ three MLP layers, in which the first two MLPs include one linear mapping layer, 1D batch normalization functional, and a ReLU activation. The last layer in $\Theta_2$ has no nonlinear activation function, while the one in $\Theta_3$ has a Sigmoid activation function. All NNs are initialized with the Xavier initialization scheme\cite{goodfellow2016regularization}. For the input CSI data, we perform antenna-wise normalization, with which the maximum of the CSI vector at each antenna element is set to unity. The following section will discuss more types of normalization.  
During the training process, we have a batch size of 256, a learning rate of $10^{-3}$ with the momentum being 0.99, and set the weight decay as $10^{-4}$. 

\textcolor{black}{The determination of the hyperparameters in the proposed MDA method is based on the manual optimization scheme, in which we manually optimize $\lambda_1$ and $\lambda_2$ from 0.0 to 1.0 with a step size being 0.1; we obtain an approximately optimal weight (denoted by $\rm{Approx. \ opt. }$) as $\lambda_1 = 0.3, \lambda_2 =0.7$. For the proposed HDA method, the related weights can be automatically optimized thanks to our theoretical analysis in Sec. III. For the weights of the task $L_{KT}$, following the work of \cite{long2018conditional}, we employ a dynamic updating strategy to update  $\lambda_3$ as $\lambda_3 \leftarrow \frac{2}{{1 + {e^{ - 10 \kappa}}}} - 1,$ where $\kappa$ starts from zero and increases with the iterative training process, making $\lambda_3$ gradually increase from 0 to 1. Lastly, we performed the network training for the weight in the auxiliary task $L_{WR}$ with  $\lambda_4$ increased from 0.01 to 0.1 with a step size of 0.01. Eventually, the optimal value of $\lambda_4$ in practice is 0.05. } 

\subsubsection{On the Communication System and the Dataset}

\textcolor{black}{The case studies are based on the Remcom 3D Ray-tracing dataset generated in the recently released dynamic outdoor scenario from the DeepMIMO website (https://deepmimo.net/scenarios/o2-scenario/). 
We focus on locating the UE(s) in a street canyon, especially the sidewalk of the main street. The main road has four lanes for mobile vehicles with different moving speeds, which change their positions for each data collection time. In this dataset, the minimum grid size (GS) for data collection is 0.2m. We use a GS of 0.8m by considering the trade-off between localization performance and data collection labor. It is noted that the vehicles have different sizes (e.g., sedans, trucks) and change their positions for each captured scene (channel realization). 
In the default setting, the UE can communicate with the BS with an $8\times8$ planar array antenna at 3.5 GHz carrier frequency with the bandwidth being 100 MHz. The maximum number of MPCs is 25. Tab. \ref{systemPara} summarizes default settings and change notes for communication system parameters in each case study. }

\textcolor{black}{Let $ {\bf{X}} \in {R^{L \times \Gamma \times M \times N}}$ and ${\bf{z}} \in R^{L \times \Gamma \times 4}$ be the collected CSI (or its variants) and related labels (3D coordinates and propagation condition indicators), respectively. $L$,  $M$, and $N$ are the number of grid points, transmit antennas, and subcarriers. $\Gamma$ is the number of measured channel realizations at each location. We collect $\Gamma=120$ scenes of data in the following experiments (''scene" here refers to a specific realization of moving blocking/scattering objects (cars) in the environment). In Wireless InSite, we can capture the CSI for all $L$ locations in one scene. For the default setting, the first fifty, next twenty, and last fifty scenes of CSI data are used for the NN training, evaluation, and test, respectively. All datasets will be organized as a mini-batch during the network training process. We will store the optimal network models that obtained the bests performance (RMSE) over the evaluation dataset and the corresponding epoch number for later convergence evaluation. In addition, we use the antenna-wise data normalization for the inputs, which are the ADP power matrix of CSI \cite{vieira2017deep, sun2018single}. 
}

\subsubsection{Evaluation Protocols}
Given some estimates ${{\bf{\hat z}}}$, the root mean square errors (RMSEs) of the localization and accuracy of propagation condition prediction are denoted by: 
{\small
\[RMSE = \frac{1}{{L \Gamma}}\sqrt {\sum\limits_{l = 1}^L {\sum\limits_{t = 1}^\Gamma {{{\left( {{x_{l,t}} - {{\hat x}_{l,t}}} \right)}^2} + {{\left( {{y_{l,t}} - {{\hat y}_{l,t}}} \right)}^2} + {{\left( {{z_{l,t}} - {{\hat z}_{l,t}}} \right)}^2}} } } \]
}
and 
$Acc = {\sum\nolimits_{l,t} {\delta \left( {{{\hat d}_{l,t}} = {d_{l,t}}} \right)} }/({{L \Gamma}}),$ where $\delta \left(  \cdot  \right)$ denotes the indicator function.

\textcolor{black}{In the following experiments, we compare the proposed methods (MDA and HDA) with three representative localization algorithms in the literature: 1) The deep DCNN based supervised learning for end-to-end location estimation method \cite{vieira2017deep}; this CNN based end-to-end localization method uses ADP as input, and the related network structure \cite{vieira2017deep} will be adopted as a backbone of feature extraction for all deep learning models. 2) the recently proposed localization method using unsupervised DA  based TL\cite{li2022long}; this method considers the distribution discrepancy caused by the dynamic environment and offers a scheme to overcome the feature space heterogeneity. 3) The scenario adaptive localization method with adversarial RDA technology, proposed in our conference paper \cite{chu2022sa}, in which the invariant representation of the network input (fingerprints) was assured with an adversarial learning scheme. Unlike the proposed methods with CR and PCP tasks, DCNN only employed the labeled data to train the CR loss. Two transfer learning methods (TL and RDA) utilized unlabeled data and novel techniques to optimize the CR task.}

\subsection{Semantic Localization v.s. Standalone Localization}
\label{Semantic1}

\begin{figure}[!t]
\centering
\subfloat[\label{pcp1} Ground truth PCPs]{ 
\includegraphics[width=0.245\textwidth]{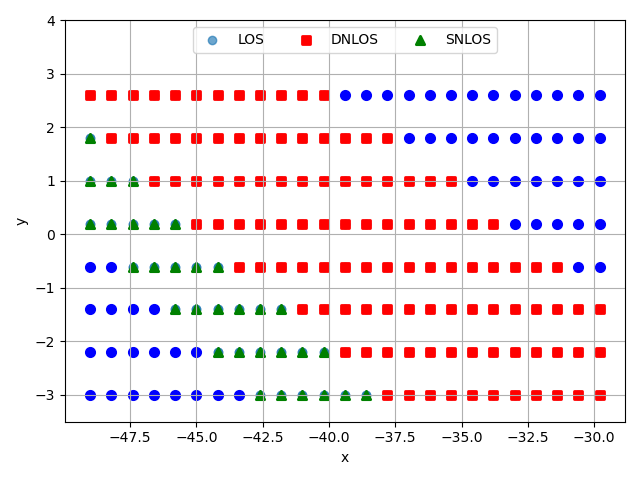}
}
\subfloat[\label{pcp2} Estimated PCPs]{
\includegraphics[width=0.245\textwidth]{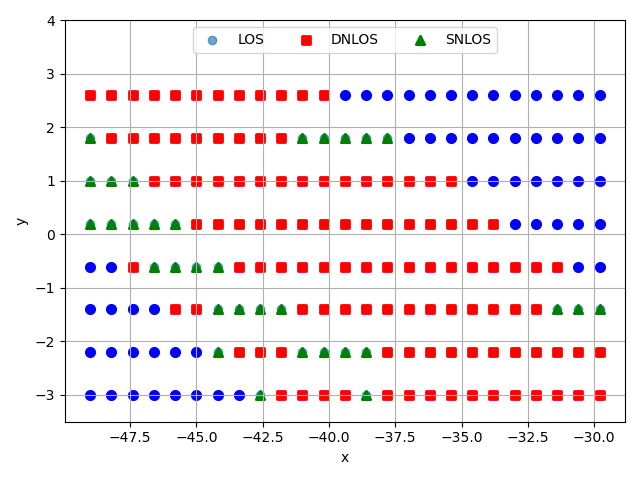}
}
\caption{The ground truth and estimated PCPs ($Environmental \ Sematics$) in a sampled area. }
\label{semantics}
\end{figure}

We first compare the semantic localization with the standalone one in terms of loss functions. The communication systems parameters are kept the same as the default setting in Tab. \ref{systemPara}, except for the available number of MPCs. To mimic the high dynamics in a street canyon, we assume that the number of MPCs ($P$ in Eq.\eqref{channel}) is a random variable, which obeys a uniform distribution $P\sim U\left( {10,25} \right)$. Tab. \ref{tabx1} comprehensively compares each loss function in \eqref{allLoss}. For obtaining the importance weights ($\lambda_1$ and $\lambda_2$) in the supervised (dominant) learning task (CR and PCP), we use three optimization strategies: 1) Unweighted scheme, where we use equal importance $\lambda_1 = \lambda_2 =0.5$; 2) Manual optimization scheme; 3) Uncertainty weighting scheme (as shown in Eq. \eqref{MLJK2}), in which the corresponding weights are denoted by the uncertainty measure and are optimized in the training process.  
Concerning the weights ($\lambda_3$ and $\lambda_4$) in KT (unsupervised learning) and WR losses, we follow the practical experience in the literature to get the corresponding optimal values.

\begin{figure*}[!ht]
\centering
\subfloat[10 Paths]{ 
\includegraphics[width=0.33\textwidth]{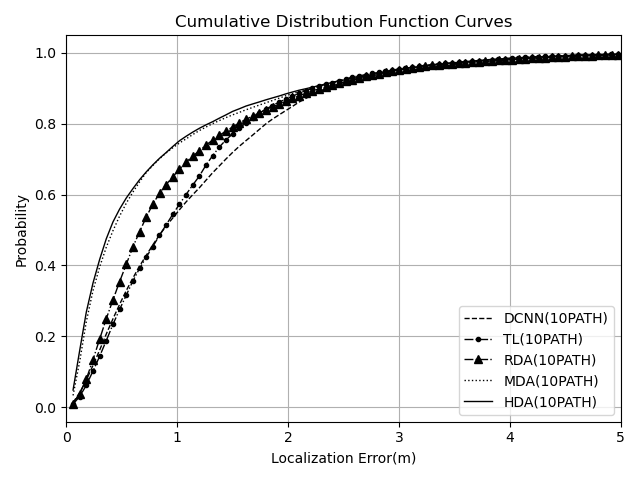}
}
\subfloat[Variable Paths]{ 
\includegraphics[width=0.33\textwidth]{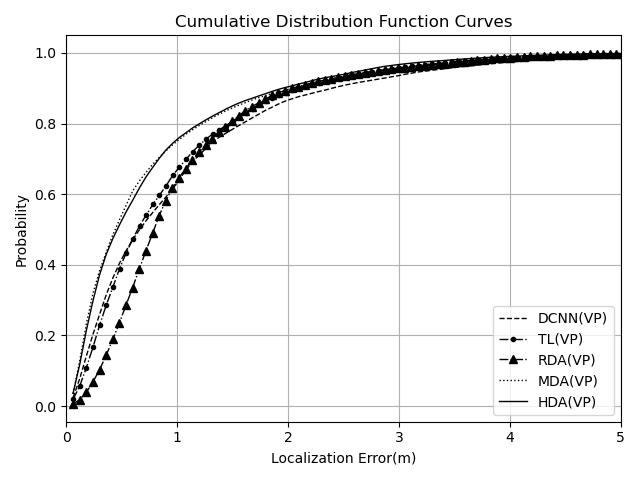}
}
\subfloat[25 Paths]{
\includegraphics[width=0.33\textwidth]{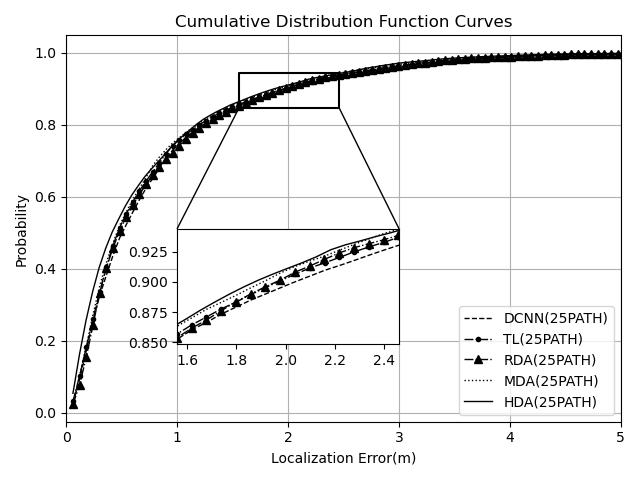}
}
\caption{The CDFs of localization errors for the three methods over different numbers of MPCs.}
\label{figMPCs}
\end{figure*}

\begin{figure}[!t]
\centering
\subfloat[20M bandwidth]{ \label{band20}
\includegraphics[width=0.245\textwidth]{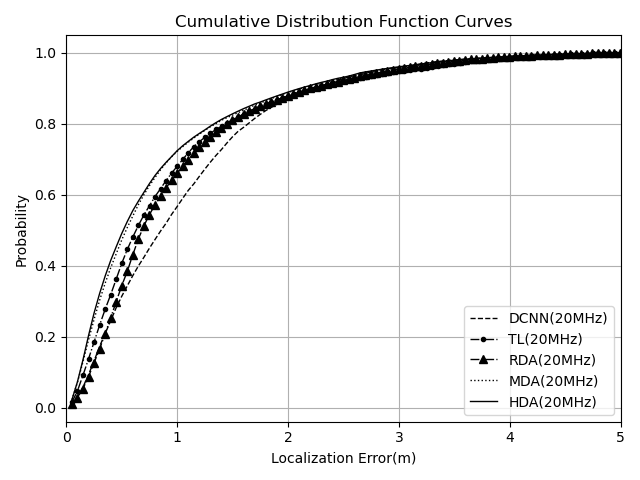}
}
\subfloat[100M bandwidth]{ \label{band100}
\includegraphics[width=0.245\textwidth]{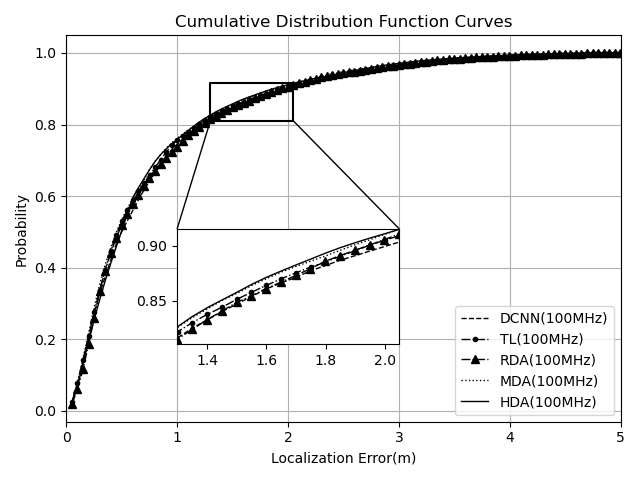}
}
\caption{The CDFs comparison of localization methods over different bandwidths}
\label{bands}
\end{figure}


The results in rows 2-4 of Tab. \ref{tabx1} show that joint training of the network with labeled and unlabeled data can boost CR and PCP performance, demonstrating the effectiveness of unsupervised DA methods. When comparing the results in the last three rows with the remainder, we find that the joint solution approach (semantic localization) consistently outperforms the localization-only solution. This can be explained by the fact that for the localization problem in the dynamic scenario, the propagation conditions between the BS and UE(s) change over time (see a sampled area in Fig.\ref{pcp1}), making the localization more difficult. On the other hand, compared to the localization-only task, the semantic localization scheme jointly estimates the coordinates of the UE(s) and the related environmental semantics (PCPs in Fig.\ref{pcp2}), which is a new dimension of the location information. This favorable information is used in the NN designs (Section \ref{SLFJT}). Moreover, the performance of the MDA method is sensitive to the importance of weights. Consequently, one might get inferior performance without optimization or fine-tuning of these weights, as verified by the results in the sixth row of Tab. \ref{tabx1}.

To get further understanding of the effect of employed environmental semantics (PCPs), we show the localization performance in a selected sampled area \ref{semantics}a, including three kinds of PCPs (LOS, DNLOS, and SNLOS). DNLOS and SNLOS are NLOS propagation conditions caused by dynamic and static objects, respectively. We compared the proposed method (MDA) in three conditions: 1)  Without using PCPs; 2) Using two kinds of PCPs: LOS and NLOS (DNLOS and SNLOS are merged together); 3) Considering three kinds of PCPs. In the investigated area, the related RMSEs of the MDA in these three conditions are 0.6774m, 0.6032m, and 0.4234m, respectively. This case study result first demonstrates that introducing environmental semantics recognition can benefit localization. More importantly, distinguishing between static and dynamic scatterers significantly improves localization performance, as indicated by DNLOS and SNLOS indications. This result indicates that the more elaborate environmental semantics, the better localization performance one can have. Together with the novel NN designs, the environmental semantics are used to reduce ambiguous features, thus enhancing the robustness of the localization algorithms against high environmental dynamics, which again explains the superiority of semantic localization.

In summary, the proposed semantic localization schemes significantly outperform standalone localization. Furthermore, integrating expert knowledge into the network optimization process improves localization accuracy. The uncertainty weighting helps to balance the different tasks in the semantic localization scheme and reduces the fine-tuning network complexity.


\subsection{Effect of Communication Systems Parameters}

This group of experiments compares all localization methods with respect to the different communication systems configurations, such as the number of MPCs (Fig. \ref{figMPCs}) and the system bandwidth (Fig. \ref{bands}).

\subsubsection{Impact of Number of MPCs}
\label{expMPCs}


We first investigate the performance of different localization methods in terms of the available number of MPCs $P$, for which we have three cases: 1) $P=10$; 2) $P$ is randomly selected from a uniform distribution $P\sim U\left( {10,25} \right)$ for each channel realization; 3) $P=25$ (Maximum in the DeepMIMO dataset). Fig. \ref{figMPCs} shows the cumulative distribution function (CDF) of comparing the three representative localization methods, and all deep learning methods can obtain better localization performance with more MPCs. Besides, it is shown in Fig. \ref{figMPCs}a that TL performs better than the supervised DCNN method as it utilizes both labeled and unlabeled data. Fig. \ref{figMPCs}b illustrates that the proposed MDA has robust performance with respect to the variation in the number of MPCs. Moreover, the multi-task learning based MDA outperforms the single-task learning based TL, showing that 1) Employing unlabeled data improves both the accuracy and robustness of localization algorithms; 2) The multiple tasks localization methods do not harm the localization performance if we can have a proper scheme for importance weight of each task, demonstrating the effectiveness of semantic localization over classical localization in challenging dynamic environments. 

\begin{figure*}[!t]
\centering
\subfloat[The method TL]{ \label{fig1xxa}
\includegraphics[width=0.33\textwidth]{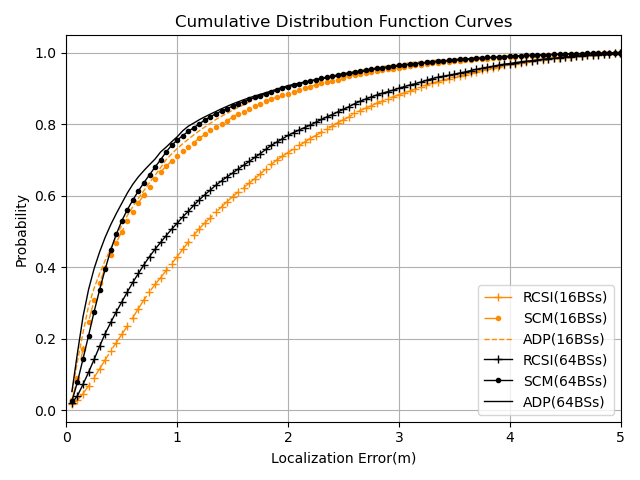}
}
\subfloat[The method RDA]{ \label{fig1xxb}
\includegraphics[width=0.33\textwidth]{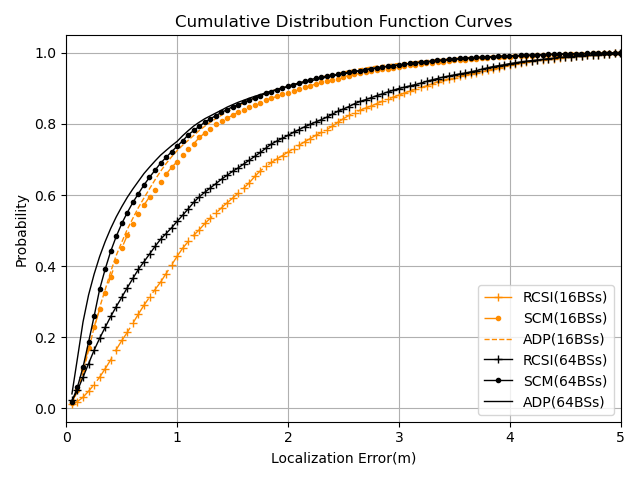}
}
\subfloat[The proposed method]{ \label{fig1xxc}
\includegraphics[width=0.33\textwidth]{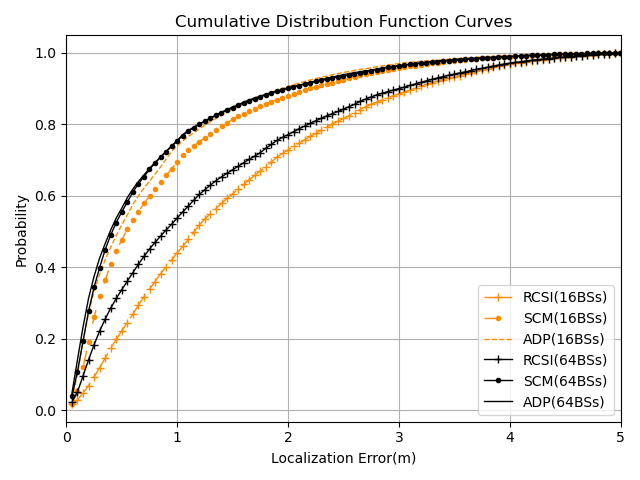}
}
\caption{The CDFs of RMSEs of different localization methods over different types of fingerprints and numbers of BS antennas.}
\label{fig1xx}
\end{figure*}

\subsubsection{The Effect of the Bandwidth}
\label{expBand}

\textcolor{black}{Next, we examine the impact of system bandwidth on localization methods. We compare all employed localization techniques using datasets bandpass filtered for 20MHz and 100MHz communication systems. This comparison helps us assess how different bandwidth settings affect the accuracy and efficiency of each method. } Other system configurations remain at their default settings (Tab. \ref{systemPara}). Fig. \ref{bands} illustrates that the system bandwidth is one of the critical parameters that affect the performance of localization algorithms. Compared to the localization method with only labeled data, the ones that jointly consider labeled and unlabeled data have significantly better performance, showing the effectiveness of adequately utilizing unlabeled data. Besides, the proposed multi-task MDA and HDA methods outperform the single-task ones, showing the superiority of semantic localization over the classical ones again.

\subsection{The Effect of the Utilization of  Data}
\label{TypesOfInputs}


\subsubsection{Types of fingerprints}

This experiment shows the effect of the numbers of transmit antennas over three types of fingerprints: 1) Real-valued channel state information (RCSI); 2) Sample covariance matrix (SCM) of CSI measurements; 3) ADP power matrix. Fig. \ref{fig1xx} shows a comparison of three DA methods in terms of CDF curves of the RMSE of localization. We used the default communication configuration to collect CSI data, except for the number of transmit antennas. It is shown in Fig. \ref{fig1xx} that the number of transmit antennas is one of the critical factors for localization methods. The ADP based localization methods obtain superior performance for both small and large numbers of transmit antennas, which can be explained by the fact that ADP preserves both the spatial and temporal ingredients of the CSI measurements while other employed fingerprints do not. Furthermore, for the 64-transmit-antenna communication system, the multi-task MDA localization with the SCM  based fingerprints can obtain comparable performance based on the ADP fingerprints, while other single-task localization methods cannot, showing the efficacy of semantic localization over the localization-only approach. 



\subsubsection{The Effect of Data Normalization} 
 \label{expNorm}

\begin{figure}[!t]
	\centering
	\includegraphics[width=0.475\textwidth]{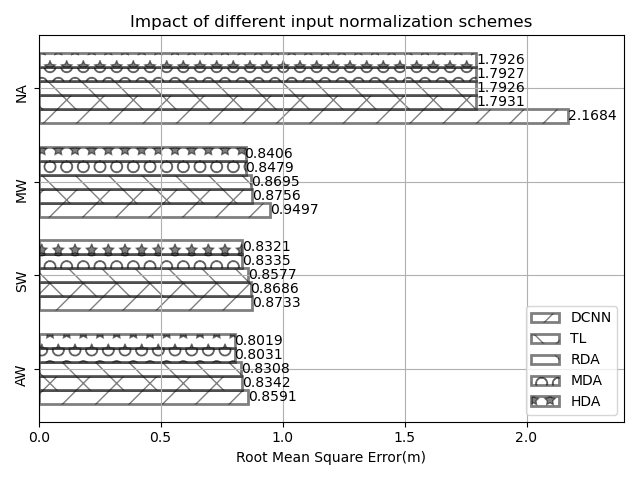}
	\caption{A comparison of different localization methods over different input normalization schemes.}
 \label{dataNorm}
\end{figure}

This experiment investigates the effect of four input data normalization schemes: 1) Antenna-wise (AW); 2) Sub-carrier wise (SW); 3) Matrix-wise (MW); 4) Normalization Absent (NA). The CSI data were collected under default communication configurations, as shown in Tab. \ref{systemPara}. We used the RMSE as the performance measure and ADP matrices as the inputs. Fig. \ref{dataNorm} shows that data normalization is vital in all investigated localization methods. The localization methods with input data normalization attain at least two times less error than those without normalization. Furthermore, the AW scheme gives the best RMSE performance and slightly outperforms the SW one. Besides, these two data normalization strategies beat the MW one, showing the necessity of preserving the individualism of each antenna or subcarrier.  

\subsection{On the Convergence and Complexity}
\label{convergence}

\textcolor{black}{The complexities of deep learning based localization methods \cite{deng2014deep, goodfellow2016regularization} are highly related to the convergence in the training stage, the model size, and the evaluation efficiency at the test stage. As a result, Tab. \ref{tabComp} reports the epoch (${\rm Epoch}^{opt}$) that the best evaluation performance happened during a total number of $2000$ epochs training. Besides, at the test stage, we utilize the floating point operations per second (FLOPs) and the number of parameters (Para.) in the networks, which can be measured by the $thop$ package from the Pytorch library. In Tab. \ref{tabComp}, we recorded the epochs to get the best performance over the validation dataset for all compared methods in a communication system with different numbers of MPCs. It is shown in Tab. \ref{tabComp} that the deep learning models consume more time (${\rm Epoch}^{opt}$) to converge when we have a more significant number of MPCs. This is reasonable as the available MPCs indicate complex wireless propagation conditions. Besides, the classical DCNN requires the least time to converge among all compared methods as it does not take advantage of unlabeled data. Moreover, the proposed MDA method requires the most extensive training time. In contrast, the proposed HDA is more efficient than MDA and performs similarly to the state-of-the-art DA based localization methods, such as RDA and TL, proving the necessity and superiority of the novel designs in HDA. }



In summary, for space and evaluation efficiency consideration (Para. and FLOPS), Tab. \ref{tabComp} shows that the proposed MDA method requires more training to achieve their best performance for the joint task as they contain novel yet complicated designs in the knowledge transfer module. On the other hand, the proposed HDA is more efficient than the MDA and exists in state-of-the-art methods such as RDA. We note that the runtime (FLOPs) of the proposed MDA method increased by $0.74 \%$ compared to existing methods (TL [53]). These relatively small relative increases in cost are rewarded by a decrease of the localization error on the order of $30 \%$, as verified in Tab. \ref{tabx1}. Regarding test performance (Para. and FLOPs), all deep learning localization methods have good complexity for many applications. The proposed methods obtain superior localization performance while consuming more memory budget and FLOPs.         

\begin{table}[!t]
\centering
\small
	\begin{tabular}{|l|l|l|ll|}
		\hline
		\multirow{2}{*}{Methods} & \multirow{2}{*}{\begin{tabular}[c]{@{}l@{}}MPCs \\  Number \end{tabular} } & \multirow{2}{*}{Epoch$^{opt}$} & \multicolumn{2}{l|}{Test}                                                 \\ \cline{4-5} 
		&                                   &                                      & \multicolumn{1}{l|}{Para.}                     & FLOPs                    \\ \hline
		\multirow{3}{*}{DCNN}    & 10 Paths                              & 265                                  & \multicolumn{1}{l|}{\multirow{3}{*}{207.378K}} & \multirow{3}{*}{8.801G}  \\ \cline{2-3}
		& VP                               & {\bf 958}                                  & \multicolumn{1}{l|}{}                          &                          \\ \cline{2-3}
		& 25 Paths                             & 926                                  & \multicolumn{1}{l|}{}                          &                          \\ \hline
		\multirow{3}{*}{TL}     & 10 Paths                                & 530                                  & \multicolumn{1}{l|}{\multirow{3}{*}{789.642K}} & \multirow{3}{*}{17.699G} \\ \cline{2-3}
		& VP                              & 1383                                  & \multicolumn{1}{l|}{}                          &                          \\ \cline{2-3}
		& 25 Paths                             & {\bf 1642}                                  & \multicolumn{1}{l|}{}                          &                          \\ \hline
		\multirow{3}{*}{RDA}     & 10 Paths                               & 443                                  & \multicolumn{1}{l|}{\multirow{3}{*}{273.556K}} & \multirow{3}{*}{8.818G}  \\ \cline{2-3}
		& VP                            & {\bf 1075}                                  & \multicolumn{1}{l|}{}                          &                          \\ \cline{2-3}
		& 25 Paths                              & 939                                  & \multicolumn{1}{l|}{}                          &                          \\ \hline
		\multirow{3}{*}{MDA}     & 10 Paths                                & 1112                                 & \multicolumn{1}{l|}{\multirow{3}{*}{1.301M}}   & \multirow{3}{*}{17.830G} \\ \cline{2-3}
		& VP                             & 1604                                  & \multicolumn{1}{l|}{}                          &                          \\ \cline{2-3}
		& 25 Paths                             & {\bf 1745}                                  & \multicolumn{1}{l|}{}                          &                          \\ \hline
		\multirow{3}{*}{HDA}     & 10 Paths                               & 631                                  & \multicolumn{1}{l|}{\multirow{3}{*}{1.301M}}   & \multirow{3}{*}{17.830G} \\ \cline{2-3}
		& VP                              & {\bf 1196}                                  & \multicolumn{1}{l|}{}                          &                          \\ \cline{2-3}
		& 25 Paths                              & 1023                                  & \multicolumn{1}{l|}{}                          &                          \\ 
    \hline
	\end{tabular} 
\caption{Complexity comparisons of all employed localization methods. \label{tabComp} }
\end{table}

\textcolor{black}{Here, we also highlight the computational burden associated with manually optimizing the four weights in the MDA method. This burden is significantly influenced by the step size chosen for the selection range. For instance, if we manually adjust $\lambda_1$ from 0.1 to 1 with a step size of 0.1, it would necessitate 10 additional training sessions to identify the optimal value for $\lambda_1$. The same procedure applies to the optimization of $\lambda_2$, $\lambda_3$, and $\lambda_4$. It is important to note that these additional computational demands for manually determining the optimal values of the weights are incurred solely during the training phase.}

\section{Conclusions}
\label{conclusions}

This paper investigated the challenging localization problem in an outdoor GPS-denied area with high dynamics. Compared to the classical localization setting, we first formulated the problem for semantic localization, taking advantage of fundamental propagation physics and improving the robustness against the high environmental dynamics. Moreover, we proposed two DA based semantic localization algorithms: 1) The proposed MDA localization method integrates the wireless expert knowledge (scenario adaptive supervised learning and knowledge transfer) into the NN design and optimization; 2) The proposed HDA localization algorithm improves the efficiency of the MDA by introducing the importance weights uncertainty modeling and the related maximum likelihood solution. Lastly, we conducted comprehensive experiments with a 3D ray tracing dataset. The case study results showed that 1) Compared to classical settings, the semantic localization scheme gives better and more robust localization performance in a dynamic environment; 2) The component settings of the localization system, including communication system configurations, the NN optimization schemes, and the utilization of input data, have a high impact on the localization results. The proposed localization methods offer more robust and precise results than existing ones over a wide range of experimental settings, demonstrating the benefits of applying the proposed algorithms for the accurate localizations of pedestrian/self-driving vehicles.  

\textcolor{black}{Throughout the case studies, it is found that environmental dynamics significantly affect the deep learning models. A proper scheme, such as the one in the proposed method, can be adopted to reduce the effect. 
\textcolor{black}{While integrating semantic identification does not always confer an advantage, a suitable method like the proposed HDA can alleviate such challenges. Additionally, we offer a theoretical framework for uncertainty modeling, facilitating the automatic refinement of importance weights in each supervised loss function. Leveraging the global convergence of NNs with appropriate bounds for the weights of all loss functions would be an important future endeavor.}
Lastly, verification of results based on actual measurements with a (hardware) channel sounder will be reported in future work.}

\ifCLASSOPTIONcaptionsoff
  \newpage
\fi



%

\bibliographystyle{IEEEtran}
\bibliography{reference}

%

\begin{IEEEbiography}[{\includegraphics[width=1in,height=1.25in,clip,keepaspectratio]{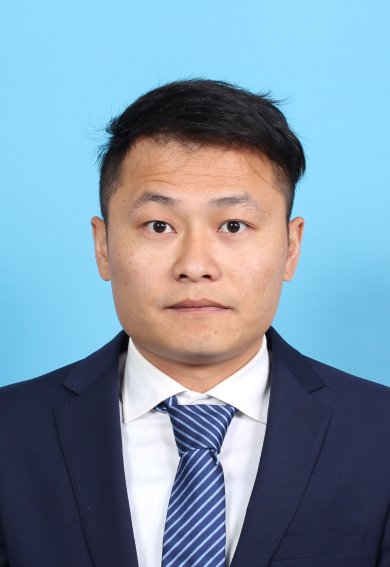}}]{Lei Chu}
		(Senior Member, IEEE) received the
		Ph.D. degree from Shanghai Jiao Tong University
		(SJTU) in 2020. He has been a full-time
		Research Scholar with the University of Southern
		California, Los Angeles, CA, USA, since 2021. Before that,
		he was a research fellow with the School of Electronics, Information, and Electrical Engineering,
		SJTU. 
  
  His current research focuses on combining learning theory and data physics with neural network optimization to advance wireless communications and intelligent sensing applications. He has contributed
		to three book chapters, authored over fifty papers in refereed
		journals/conferences, and issued many patents.  He serves as a regular reviewer for dozens of IEEE journals and conferences. He delivered a tutorial at IEEE GLOBECOM 2022 titled "Scalable, Accurate, and Privacy-Preserving Localization in B5G Wireless Networks." He received the Outstanding Master's Thesis Award in 2015 and the Outstanding Ph.D. Graduate Award in 2020. He serves as an associate editor for \textit{IEEE Open Journal of Vehicular Technology, IET Electrical Systems in Transportation, and Journal of Artificial Intelligence and Big Data}. 
	\end{IEEEbiography}

\begin{IEEEbiographynophoto}{Abdullah Alghafis}
Dr. Alghafis's biography is not available in the current version.
\end{IEEEbiographynophoto}

\begin{IEEEbiography}[{\includegraphics[width=1.05in,height=1.25in,clip,keepaspectratio]{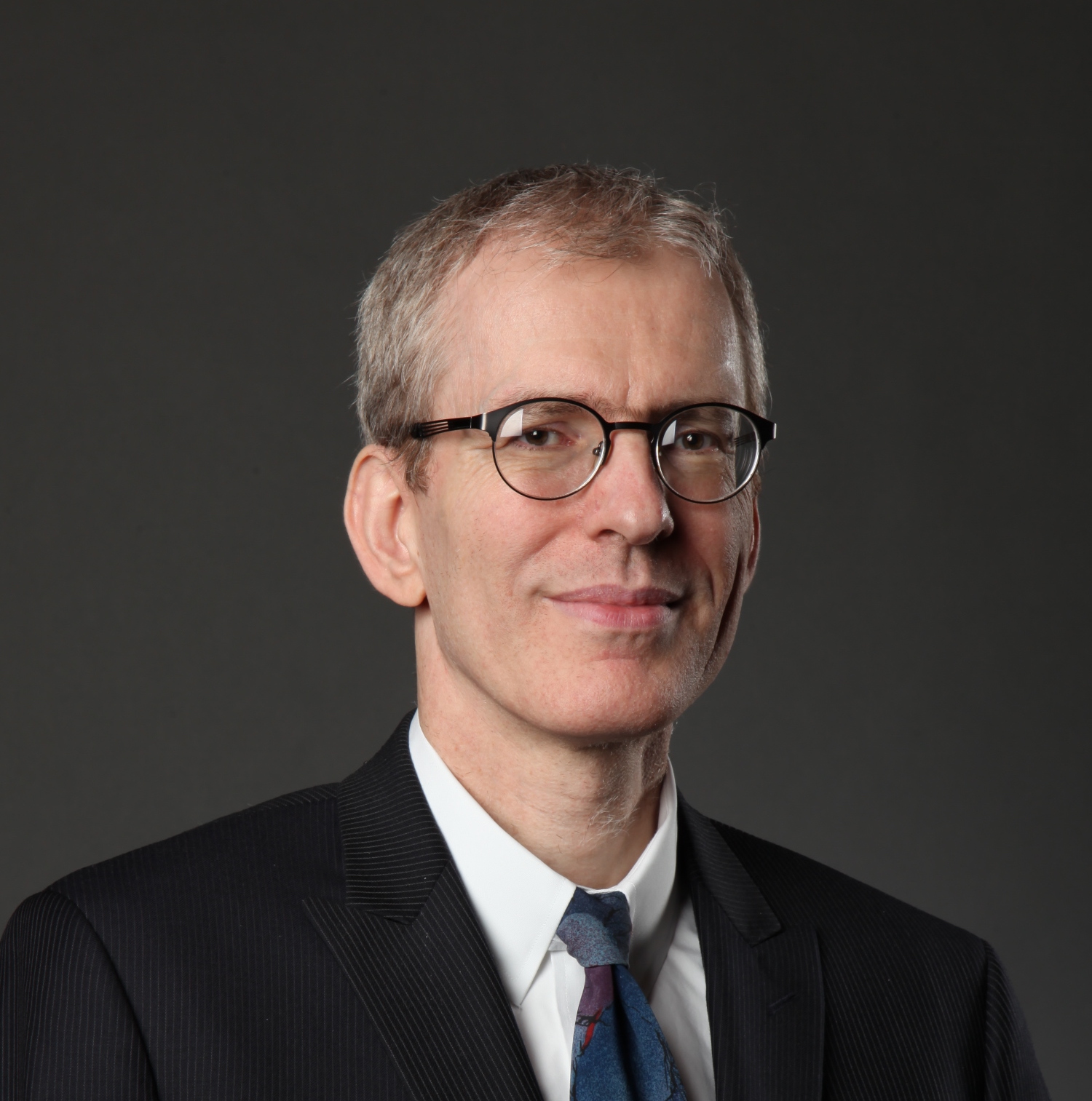}}]{Andreas F. Molisch}
		(Fellow, IEEE)
received his degrees (Dipl.Ing. 1990, PhD 1994, Habilitation 1999) from the Technical University Vienna, Austria. He spent the next 10 years in industry, at FTW, AT\&T (Bell) Laboratories, and Mitsubishi Electric Research Labs (where he rose to Chief Wireless Standards Architect). In 2009 he joined the University of Southern California (USC) in Los Angeles, CA, as Professor, and founded the Wireless Devices and Systems (WiDeS) group. In 2017, he was appointed to the Solomon Golomb – Andrew and Erna Viterbi Chair. 

His research interests revolve around wireless propagation channels, wireless systems design, and their interaction. Recently, his main interests have been wireless channel measurement and modeling for 5G and beyond 5G systems, joint communication-caching-computation, hybrid beamforming, UWB/TOA based localization, and novel modulation/multiple access methods. Overall, he has published 5 books (among them the textbook “Wireless Communications”, third edition in 2023), 22 book chapters, >300 journal papers, and >400 conference papers. He is also the inventor of 70 granted (and more than 10 pending) patents, and co-author of some 70 standards contributions. His work has been cited more than 68,000 times, his h-index is 110, and he is a Clarivate Highly Cited Researcher. 

Dr. Molisch has been an Editor of a number of journals and special issues, General Chair, Technical Program Committee Chair, or Symposium Chair of multiple international conferences, as well as Chairperson of various international standardization groups. He is a Fellow of the National Academy of Inventors, Fellow of the AAAS, Fellow of the IEEE, Fellow of the IET, an IEEE Distinguished Lecturer, and a member of the Austrian Academy of Sciences. He has received numerous awards, among them the IET Achievement Medal, the Technical Achievement Awards of IEEE Vehicular Technology Society (Evans Avant-Garde Award) and the IEEE Communications Society (Edwin Howard Armstrong Award), and the Technical Field Award of the IEEE for Communications, the Eric Sumner Award.

\end{IEEEbiography}







\end{document}